\PassOptionsToPackage{pdfpagelabels=false}{hyperref}
\pdfoutput=1 

\documentclass[fleqn,usenatbib]{mnras}

\usepackage{newtxtext,newtxmath}

\usepackage[T1]{fontenc}
\usepackage{ae,aecompl}

\usepackage{amsmath}
\usepackage{amssymb}
\usepackage{aas_macros}
\usepackage{rotating}
\usepackage{xspace}
\usepackage{booktabs}
\usepackage{multirow}
\usepackage{xcolor}
\usepackage{listings}

\usepackage[newfloat, frozencache]{minted} 
\definecolor{vargreen}{HTML}{00BB00}
\definecolor{funccolor}{HTML}{AA22FF}
\SetupFloatingEnvironment{listing}{name=Code,placement=htp}

\usepackage{url}
\usepackage{textcomp}
\usepackage{gensymb}
\usepackage{printlen}
\usepackage[export]{adjustbox}[2011/08/13]
\usepackage{esvect} 

\usepackage{tikz}
\usepackage{pgfplots}
\pgfplotsset{compat = newest}
\usetikzlibrary{calc, quotes, angles, arrows, fit, arrows.meta, decorations.markings}

\newcommand{\MarkRightAngle}[4][.3cm]
{\coordinate (tempa) at ($(#3)!#1!(#2)$);
 \coordinate (tempb) at ($(#3)!#1!(#4)$);
 \coordinate (tempc) at ($(tempa)!0.5!(tempb)$);
 \draw [very thick](tempa) -- ($(#3)!2!(tempc)$) -- (tempb);
}

\newcommand{\hMsun}{\ensuremath{h^{-1}M_{\odot}}}
\newcommand{\hMpc}{\ensuremath{{h^{-1}\mathrm{Mpc}}\xspace}}
\newcommand{\rmax}{\texorpdfstring{\ensuremath{{{\mathcal{R}}_\mathrm{max}}}\xspace}{rmax}}
\newcommand{\rmin}{\texorpdfstring{\ensuremath{{{\mathcal{R}}_\mathrm{min}}}\xspace}{rmin}}
\newcommand{\thetamax}{\texorpdfstring{\ensuremath{\theta_\mathrm {max}}\xspace}{thetamax}}
\newcommand\vvec[1]{\ensuremath{{\mathbf{#1}}}}
\newcommand\norm[1]{\left\lVert#1\right\rVert}
\newcommand{\pimax}{\texorpdfstring{\ensuremath{{\pi_\mathrm{max}}}\xspace}{pimax}}
\newcommand{\pimaxsqr}{\texorpdfstring{\ensuremath{{\pi^{2}_\mathrm{max}}}\xspace}{pimax^2}}
\newcommand{\rmaxcubed}{\ensuremath{{\mathcal{R}^{3}_\mathrm{max}}}\xspace}
\newcommand{\rmaxsqr}{\ensuremath{{\mathcal{R}^{2}_\mathrm{max}}}\xspace}
\newcommand{\rsep}{\ensuremath{{{\mathcal{R}}_\mathrm{sep}}}\xspace}
\newcommand{\lbox}{\ensuremath{\mathcal{L}}\xspace}

\newcommand{\npart}{\ensuremath{\mathcal{N}}\xspace}
\newcommand{\npartsqr}{\ensuremath{\mathcal{N}^2}\xspace}
\newcommand{\onpart}{\ensuremath{\mathcal{O}(\npart)}\xspace}
\newcommand{\onlogn}{\ensuremath{\mathcal{O}(\npart\log\npart)}\xspace}
\newcommand{\onpartsqr}{\ensuremath{\mathcal{O}(\npartsqr)}\xspace}

\newcommand{\corrfunc}{\texttt{Corrfunc}\xspace}
\newcommand{\lcdm}{\ensuremath{\mathrm{\Lambda CDM}}\xspace}
\newcommand{\xir}{\ensuremath{{DD(r)}}\xspace}
\newcommand{\xiofr}{\ensuremath{{\xi(r)}}\xspace}
\newcommand{\wprp}{\ensuremath{{w_p(r_p)}}\xspace}
\newcommand{\xirppi}{\ensuremath{{\xi(r_p,\pi)}}\xspace}

\newcommand{\ddrppi}{\ensuremath{{DD(r_p,\pi)}}\xspace}
\newcommand{\drrppi}{\ensuremath{{DR(r_p,\pi)}}\xspace}
\newcommand{\wtheta}{\ensuremath{{\omega(\theta)}}\xspace}
\newcommand{\ddtheta}{\ensuremath{{DD(\theta)}}\xspace}
\newcommand{\drtheta}{\ensuremath{{DR(\theta)}}\xspace}
\newcommand{\RA}{{\small {RA}}\xspace}
\newcommand{\DEC}{{\small {DEC}}\xspace}
\newcommand{\mcmc}{{\small {MCMC}}\xspace}
\newcommand{\hpc}{{\small {HPC}}\xspace}

\newcommand{\openmp}{\texttt{OpenMP}\xspace}
\newcommand{\avxft}{\texttt{AVX512}\xspace} 
\newcommand{\avxftf}{\texttt{AVX512F}\xspace} 
\newcommand{\avx}{\texttt{AVX}\xspace}
\newcommand{\sse}{\texttt{SSE}\xspace}
\newcommand{\fallback}{\texttt{Fallback}\xspace}
\newcommand{\simd}{\texttt{SIMD}\xspace}
\newcommand{\simdlen}{\texttt{SIMDLEN}\xspace}
\newcommand{\RAM}{\texttt{RAM}\xspace}

\newcommand{\halotools}{\texttt{halotools}\xspace}
\newcommand{\kdcount}{\texttt{kdcount}\xspace}
\newcommand{\treecorr}{\texttt{TreeCorr}\xspace}
\newcommand{\sklearn}{\texttt{scikit-learn KDTree}\xspace}
\newcommand{\cutebox}{\texttt{CUTE\_box}\xspace}
\newcommand{\scipy}{\texttt{SciPy cKDTree}\xspace}
\newcommand{\mlpack}{\texttt{mlpack RangeSearch}\xspace}
\newcommand{\swot}{\texttt{swot}\xspace}

\newcommand{\figwidth}{0.75}

\newcommand{\eye}[4]
{   \draw[rotate around={#4:(#2,#3)}] (#2,#3) -- ++(-.5*55:#1) (#2,#3) -- ++(.5*55:#1);
    \draw (#2,#3) ++(#4+55:.75*#1) arc (#4+55:#4-55:.75*#1);
    \draw[fill=gray] (#2,#3) ++(#4+55/3:.75*#1) arc (#4+180-55:#4+180+55:.28*#1);
    \draw[fill=black] (#2,#3) ++(#4+55/3:.75*#1) arc (#4+55/3:#4-55/3:.75*#1);
}

\title[\corrfunc\ --- Blazing Fast Correlation functions]{\corrfunc\ --- A Suite of Blazing Fast Correlation Functions on the CPU}
\author[Sinha \& Garrison]{
Manodeep Sinha,$^{1, 2, 3}$\thanks{E-mail: msinha@swin.edu.au}
Lehman H.~Garrison$^{4, 5}$
\\
$^{1}$SA 101, Centre for Astrophysics \& Supercomputing, Swinburne University of Technology, 1 Alfred St., Hawthorn, VIC 3122, Australia\\
$^{2}$ARC Centre of Excellence for All Sky Astrophysics in 3 Dimensions (ASTRO 3D)\\
$^{3}$6301 Stevenson Center, Department of Physics \& Astronomy, Vanderbilt University, Nashville, TN 37235\\
$^{4}$Center for Computational Astrophysics, Flatiron Institute, 162 Fifth Ave., New York, NY 10010\\
$^{5}$Center for Astrophysics | Harvard \& Smithsonian, 60 Garden St, Cambridge, MA 02138
}

\date{Accepted XXX. Received YYY; in original form ZZZ}

\pubyear{\the\year{}}

\begin{document}
\label{firstpage}
\pagerange{\pageref{firstpage}--\pageref{lastpage}}
\maketitle

\begin{abstract}
The two-point correlation function (2PCF) is the most widely used tool for quantifying the spatial distribution of galaxies. Since the distribution of galaxies is determined by galaxy formation physics as well as the underlying cosmology, fitting an observed correlation function yields valuable insights into both. The calculation for a 2PCF involves computing pair-wise separations and consequently, the computing time scales quadratically with the number of galaxies. The next-generation galaxy surveys are slated to observe many millions of galaxies, and computing the 2PCF for such surveys would be prohibitively time-consuming. Additionally, modern modelling techniques require the 2PCF to be calculated thousands of times on simulated galaxy catalogues of {\em at least} equal size to the data and would be completely unfeasible for the next generation surveys. Thus, calculating the 2PCF forms a substantial bottleneck in improving our understanding of the fundamental physics of the universe, and we need high-performance software to compute the correlation function.
In this paper, we present \corrfunc --- a suite of highly optimised, \openmp parallel clustering codes. The improved performance of \corrfunc arises from both efficient algorithms as well as software design that suits the underlying hardware of modern CPUs. \corrfunc can compute a wide range of 2-D and 3-D correlation functions in either simulation (Cartesian) space or on-sky coordinates.  \corrfunc runs efficiently in both single- and multi-threaded modes and can compute a typical 2-point projected correlation function (\wprp) for $\sim 1$ million galaxies within a few seconds on a single thread. \corrfunc is designed to be both user-friendly and fast and is publicly available at \url{https://github.com/manodeep/Corrfunc}.
\end{abstract}


\section{Introduction}
In an \lcdm cosmology, galaxies form and evolve in dark  matter  halos.  The  spatial  distribution  of galaxies is  therefore dictated by the underlying dark  matter halos, which in turn, depends sensitively on the cosmological parameters.  Thus, the  observed  clustering  of  galaxies  contains  information about the way galaxies occupy host dark matter halos - the {\em galaxy-halo connection}, as well as the cosmological model. Therefore, by studying the clustering of galaxies, we can gain valuable information about both cosmology and the galaxy-halo connection. Consequently, there exists a wide range of clustering statistics, e.g.,  the two-point correlation function, the  void  probability  function,  the  3-point  correlation function, the pair-wise velocity dispersion, that inspect different aspects of the galaxy-halo connection or the cosmological model. In this paper, we will
focus on the most commonly used clustering statistic---the 2-point correlation function (2PCF).

The 2PCF is a powerful probe of both cosmology and structure formation. Typically, we extract cosmological information  from  the  correlation  function  at  `large'  separations, while the `small' separations also contain the imprints of the complex galaxy formation processes.
The enormous constraining power of the 2PCF can be highlighted through the myriad applications on a wide range of research questions. For example, the 2PCF has been used to constrain the cosmological model~\citep{Eisenstein+2005, tegmark2006_lrg, Blake2011_wigglez_bao, Beutler2011_BAO, Percival+2010, Anderson+2014, Hildebrant2016_kids_photoz_validation, Alam+2017}, dissecting halo clustering~\citep[e.g.,][]{Gao2005_assemblybias, wechsler_halo_assembly_bias_2006, salcedo_secondary_bias_2018}, probing the epoch of reionization~\citep[e.g.,][]{mcquinn2007_lae, ouchi2018_silverrush_lae}, validating galaxy photometric redshifts~\citep[e.g.,][]{Hildebrant2016_kids_photoz_validation, gatti2018_photoz_validation_theory}. The 2PCF is instrumental for investigating various aspects of the galaxy-halo connection -- e.g, the clustering of galaxies as a function of luminosity~\citep[e.g.,][]{Norberg2002_2df_clustering, yang_clf_2003, zehavi_luminosity_clustering_2005, conroy_sham_2006}, stellar mass~\citep[e.g.,][]{moster_sm_hm_2010,  leauthaud_stellar_mass_to_dm_2012,zu_stellar_mass_2015}, galaxy colour~\citep[e.g.,][]{zehavi_color_clustering_2011, hearin_darkside_2013}.
Thus, we can extensively probe the physics of galaxy formation and evolution with the 2PCF (see \citealt{Wechsler_Tinker_2018} for a recent overview).

We advance our understanding of the Universe through (at least) these two approaches -- i) increasing how precisely we can determine essential model parameters (e.g., determining the Hubble constant to 1\% precision) and ii) studying the relative clustering strengths within sub-samples of galaxies grouped by similar physical properties (e.g., for galaxies at fixed stellar mass, do redder galaxies live in more massive halos compared to bluer galaxies?). Both these scenarios benefit from more precise correlation functions resulting from larger galaxy samples. Current surveys like the Sloan Digital Sky Survey (SDSS) \citep{Blanton+2017} have already produced catalogues with millions of galaxies. Upcoming surveys, both photometric and spectroscopic, e.g., Dark Energy Spectroscopic Instrument (DESI) Survey~\citep{Levi+2013}, \textit{Euclid}~\citep{Laureijs+2011}, and Large Synoptic Survey Telescope~\citep{lsstSRD} will probe even larger volumes and fainter galaxies and target 10s of millions to billions of galaxies. With such a wealth of galaxy data, we can measure the galaxy density field more precisely than ever before.

Such exquisite data from existing and upcoming galaxy surveys bring their challenges. A brute-force approach to computing the 2PCF requires pairwise separations between all possible pairs of galaxies, i.e., the 2PCF has a computational complexity of \onpartsqr, where \npart is the number of input galaxies. For instance, to compute the 2PCF for $10^6$ galaxies, we would first need to compute $10^{12}$ separations. Even with the fastest computers available today, computing $10^{12}$ distances will take significant time. For tens of millions of galaxies, it will take days to weeks to compute the 2PCF with such a brute-force approach.

The computational demand becomes even more extreme when we consider modern modelling techniques like Bayesian inference within a Monte Carlo Markov Chain (\mcmc). To obtain converged parameter estimates in an \mcmc analysis, we need to generate many different realisations of the theoretical galaxy distribution corresponding to plausible combinations of parameter values. Each realisation, potentially containing $\sim$ millions of galaxies, requires a new 2PCF computation. Even if each such 2PCF calculation only takes five minutes, then repeating such a brute-force 2PCF calculation for $\gtrsim 10^5$ iterations will take $\gtrsim 1$ year. Such a timescale would rule out any attempts to reproduce the observed galaxy clustering within an \mcmc analysis. Thus, if we want to model the upcoming galaxy surveys within a Bayesian framework, a faster correlation function code is {\em critical}.

In recent years, at least two research communities\footnote{Astronomy is not the only field facing the computational challenge of correlation functions.  Techniques in computer science and molecular dynamics have been developed to solve similar problems \citep[e.g.][]{spatial_distance_histogram_performance_chen_2011}.} have developed correlation function codes: the high-performance computing (\hpc) community \citep[e.g.][]{Chhugani2012,mlpack} and the astronomical community \citep[e.g.,][also, see the code comparison in \S~\ref{sec:code_comparison} for more examples]{treecorr, cute_alonso_2012, swot}.  The codes from the \hpc community tend to focus on the technical challenges rather than the scientific outcome. For instance, common astronomy use-cases like the angular correlation function, are rarely addressed by the \hpc codes. On the other hand, correlation function codes written by astronomers tend to be slower and problem-specific. \corrfunc is designed to fill this gap --- a high-performance, well-tested, well-documented, flexible, open-source code for computing most kinds of correlation functions straight out of the box.

The paper is structured in the following manner - in \S~\ref{sec:background}, we will discuss the basic implementation of a correlation function and provide a broad overview of the \corrfunc software package, in \S~\ref{sec:cpu_primer} we will discuss the aspects of computing hardware relevant for the design of a high-performance code, in \S~\ref{sec:corrfunc_methods} we will discuss the optimisations implemented in \corrfunc\footnote{In this paper we describe \corrfunc \texttt{v2.0.0}.}, the performance and scaling in \S~\ref{sec:benchmarks}, compare the runtime performance of \corrfunc with other existing open-source correlation in \S~\ref{sec:code_comparison}. We will discuss the shortcomings and future directions for \corrfunc in \S~\ref{sec:discussion} and then conclude in \S~\ref{sec:conclusion}.

\section{Background}\label{sec:background}
A correlation function is a measure of the excess probability of finding a pair of galaxies separated by spatial scale $r$ or angular scale $\theta$.  The classic spatial 2PCF, $\xi(r)$, and the angular 2PCF, $\omega(\theta)$, are defined as:
\begin{align}
\begin{split}
dP &= n_g(r) \left [1 + \xi(r)\right] dV, \\
dP &= \mathcal{N}_g(\theta) \left[1 + \omega(\theta)\right] d\Omega,
\end{split}
\end{align}
where $dP$ is the excess probability, $n_g(r)$ and $\mathcal{N}_g(\theta)$ are the mean densities of galaxies at the given separation scale, and $dV$ and $d\Omega$ are the differential volume and solid angle elements.

Regardless of the correlation function type, the fundamental operation to obtain a correlation function is to compute separations between pairs of galaxies. Therefore, the positions of galaxies are a required input to compute a correlation function. While the exact positions of simulated galaxies are directly available, the positions of observed galaxies are derived from the observed redshift. The observed redshift of a galaxy is a combination of the cosmological recession velocity and the line-of-sight projection of the galaxies' peculiar velocity. Since we can
not disentangle the two contributions, we can not infer the actual position of observed galaxies and thus can not measure $\xiofr$. However, we know that the peculiar velocity component spreads out galaxies along the line-of-sight. If we measure the correlation function as a two-dimensional histogram, $\xirppi$, i.e., count galaxy pairs as a function of both the projected separation ($r_p$) and line-of-sight separation ($\pi$), then we can account for the effect peculiar velocity by integrating pairs along the line of sight. The resultant correlation function is called the projected two-point correlation function --- $\wprp$ --- and is defined as:
\begin{align}
\begin{split}
\wprp &= \int_{-\infty}^\infty \xi(r_p, \pi) d\pi, \\
      &\approx 2\times\int_0^\pimax \xi(r_p, \pi) d\pi.
\end{split}
\end{align}
Assuming isotropy, we can introduce a factor of 2 and switch the lower limit to zero.  While integrating to infinity along the line of sight is guaranteed to remove all effects of the peculiar velocity, in practice, since galaxy surveys do not extend to infinity, we need a finite upper limit on the integral (\pimax). \pimax needs to be sufficiently large to nullify the effect of peculiar velocities, while not too large to create artificial edge effects from the survey boundary. Typically \pimax is chosen to be in the range $40-80$ Mpc, with the exact value of \pimax determined as appropriate for the underlying galaxy survey~\citep[see][for a discussion of the errors from a finite \pimax]{vandenbosch_pimax_2013}.

\subsection{How to compute a correlation function}
A correlation function is defined as the excess clustering of a target distribution of galaxies over a random distribution. Thus, to measure the correlation function, we require at least two terms -- one term that measures the distribution of galaxies (the ``data" term) and another term that measures the distribution of a random distribution with the same number-density as the galaxies (the ``randoms" term). The randoms term helps to both quantify the excess clustering and correctly account for the survey edges and the survey incompleteness. The simplest estimator for the correlation function, the natural estimator\footnote{The commonly used ``Landy-Szalay estimator"~\citep{landy_szalay_1993} also uses an additional ``DR" term for a better estimate of the correlation function.}, is written as: $1 + \xi(r) = DD(r)/RR(r)$, where $DD(r)$ and $RR(r)$ indicate the number of ``galaxy-galaxy" and ``random-random" pairs respectively, with the pair-separation in the range $[r-dr/2, r + dr/2)$. From the computational viewpoint, calculating a correlation function requires computing the pair-wise separations between pairs of points and then creating a (possibly weighted) histogram out of the computed pair separations.
For two data-sets with \texttt{N1} and \texttt{N2} points, the simplest possible (brute-force) implementation for a pair-counting code is shown in Code~\ref{code:naivecorr}.
\begin{listing}
\caption{\small{Naive C code for a correlation function}}
\begin{minted}[escapeinside=||]{C}
for(int i=0;i<N1;i++){
  for(int j=0;j<N2;j++){
    double dist = |\textcolor{funccolor}{distance\_metric}|(i, j);
    if(dist < mindist || dist >= maxdist){
      continue;
    }

    int ibin = |\textcolor{funccolor}{dist\_to\_bin\_index}|(dist);
    numpairs[ibin]++;
    weight[ibin] += |\textcolor{funccolor}{weight\_func}|(i, j);
  }
}
\end{minted}
\label{code:naivecorr}
\end{listing}

Examining the code snippet, we can see that {\em at most} three functions, viz., \texttt{distance\_metric}, \texttt{dist\_to\_bin\_index} and \texttt{weight\_func}, are necessary to fully describe an arbitrary correlation function. These three functions perform the following tasks:
\begin{itemize}
\item \texttt{distance\_metric} --- Quantifies the attributes of the individual points (in the pair) into a separation. For 3-D Euclidean geometries, this mapping is simply $d_{ij}^2 = (x_i - x_j)^2 + (y_i - y_j)^2 + (z_i - z_j)^2$.
\item \texttt{dist\_to\_bin\_index} --- Converts the computed separation into a bin index for the corresponding correlation function. Traditionally, logarithmic bins are used for $\xi(r)$ and $\wprp$; however, binning could be in two/multiple dimensions with independent logarithmic/linear choices in each dimension.
\item \texttt{weight\_func} --- Quantifies the contribution from a given pair of points. Not always required but allows for more complex selection of the pair as well as accounting for survey incompleteness. \texttt{weight\_func} is usually the product of the weights for each point in the pair (i.e., $w_i \times w_j$) but arbitrary kinds of weighting schemes can be specified for a correlation function\footnote{See \url{https://halotools.readthedocs.io/en/latest/api/halotools.mock_observables.marked_tpcf.html\#halotools.mock_observables.marked_tpcf} for examples of different weighting functions.}.
\end{itemize}

\corrfunc was designed to accommodate different combinations of  \texttt{distance\_metric}, \texttt{dist\_to\_bin\_index} and \texttt{weight\_func}. In the following subsection, we will go over broad design goals of \corrfunc and the various pair-counters available in \corrfunc.

\subsection{Package Summary}
\corrfunc is written primarily in \texttt{C} and comes with convenient
\texttt{Python 2} and \texttt{Python 3} wrappers for most clustering statistics. Since the \corrfunc code-base is updated frequently, this paper refers to \corrfunc \texttt{v2.0.0}. In the text, we have also noted where the latest version of \corrfunc (\texttt{v2.3.0}) differs from \texttt{v2.0.0}. The primary design goals for \corrfunc are the following:
\begin{itemize}
\item Correctness --- \corrfunc has a base set of correct outputs for every statistic generated either through slow, brute-force methods or independent, external codes (see \S~\ref{sec:code_comparison}). Within \corrfunc, every clustering statistic has at least one automated test case that requires reproducing the ``known-correct" number of pairs exactly.
\item High Performance --- Performance is an overarching goal for \corrfunc.  \corrfunc is parallelised for shared-memory systems via \openmp, and the most compute-intensive parts of the code have high-performance kernels written for a range of CPU architectures.
\item Portability --- \corrfunc is written in \texttt{ISO/IEC 9899:1999} compliant \texttt{C}. All hardware-specific instructions are protected via compile-time constant definitions in the source code.
\item Flexibility --- Every clustering statistic in \corrfunc can be accessed either through the \corrfunc API call (e.g., via \texttt{python} as well as the associated static library) or as a command-line executable. From each of these interfaces, minimal coding is required to implement arbitrary weighting schemes for particle pairs\footnote{The documentation on how to implement arbitrary weights within \corrfunc is here -- \url{https://corrfunc.readthedocs.io/en/master/modules/custom_weighting.html}.}.
\end{itemize}

\corrfunc is designed for the two astronomical use-cases for calculating correlation functions involving -- (i) simulated galaxies with positions in Cartesian coordinates and (ii) observed galaxies with positions in spherical coordinates. ``Mock galaxies", corresponding to simulated galaxies that have been projected on to the sky\footnote{There might be additional layers of observational realism added in to make the mock galaxies more closely resemble the observed galaxies}, can be treated as observed galaxies.  \corrfunc contains four and two correlation function routines for simulated and observed galaxies, respectively. The routines corresponding to the simulated galaxies are located in the \texttt{theory} directory while the routines for the observed galaxies are located in the \texttt{mocks} directory. The primary difference between the \texttt{theory} and \texttt{mocks} routines is the definition of the line-of-sight distance -- we assume the plane-parallel approximation for the \texttt{theory} routines and the \cite{fisher_1994_rp_pi} convention for the \texttt{mocks} routines (see Appendix~\ref{sec:appendix_separations} for details). In Table~\ref{table:corrfuncs}, we list the available routines and their expected inputs. These distinct correlation functions target the commonly used conventions for \texttt{distance\_metric} and \texttt{dist\_to\_bin\_index}.

\begin{table*}
\centering
\caption{Available correlation function routines with \corrfunc \texttt{v2.0.0}. The separation metrics used in each of the correlation functions are outlined in Appendix~\ref{sec:appendix_separations}. The inputs for the \texttt{theory} correlation functions are \texttt{X, Y, Z} -- typically Cartesian co-moving positions from simulations. The inputs for the \texttt{mocks} are Right Ascension (\texttt{RA}) and Declination (\texttt{DEC}), within the range $[0\degree,360\degree]$ and $[-90\degree, 90\degree]$ respectively. The \texttt{DDrppi\_mocks} requires an additional input \texttt{CZ} - the product of the redshift and the speed of light, expected to be in units of \texttt{km/s}. }
\label{table:corrfuncs}
\begin{adjustbox}{max width=\linewidth}
\begin{tabular}{llllll}
\toprule
\multirow{2}{*}{\textbf{Type}}                          &
\multirow{2}{*}{\textbf{Input positions}}               &
\multirow{2}{*}{\textbf{Name}}                          &
\multirow{2}{*}{\textbf{C source directory}}                      &
\multicolumn{1}{c}{\textbf{Python wrapper}}             &
\multirow{2}{*}{\textbf{Bins}}                \\
& & & & \multicolumn{1}{c}{under directory: \texttt{Corrfunc}} & \\
\midrule
\multirow{4}{*}{\textbf{Theory}} & \multirow{4}{*}{\texttt{X, Y, Z}} & $\xir$     & \texttt{theory/DD/}     & \texttt{theory/DD.py}      & $r$\\
                                 &                                   & $\ddrppi$  & \texttt{theory/DDrppi/} & \texttt{theory/DDrppi.py}  & $(r_p, \pi)$\\
                                 &                                   & $\wprp$    & \texttt{theory/wp/}     & \texttt{theory/wp.py}      & $r_p$\\
                                 &                                   & $\xiofr$   & \texttt{theory/xi/}     & \texttt{theory/xi.py}      & $r$\\
\cmidrule(l{0.5em}r{0.5em}){1-6}
\multirow{2}{*}{\textbf{Mocks}}  & \texttt{\RA, \DEC, CZ} & $\ddrppi$   & \texttt{mocks/DDrppi\_mocks/}   & \texttt{mocks/DDrppi\_mocks.py}    & $(r_p, \pi)$\\
                                 & \texttt{\RA, \DEC}     & $\ddtheta$  & \texttt{mocks/DDtheta\_mocks/}  & \texttt{mocks/DDtheta\_mocks.py}   & $\theta$\\
\bottomrule
\end{tabular}
\end{adjustbox}
\end{table*}

Additionally, one of the design goals for \corrfunc is the user experience, starting right from the installation step. Installation is designed to be free of user input by default --- compile, and link options are automatically populated, and paths to runtime dependencies are embedded into the \corrfunc shared library for \texttt{python}. We have undertaken a significant effort to ensure that \corrfunc compiles straight out of the box for the typical user, while also providing the option to the advanced user to customise their install of \corrfunc. However, in this paper, we will focus on the algorithm and the high-performance aspects of the \corrfunc package and leave the design for user-experience for a separate occasion.

Before we delve into the \corrfunc package design and optimisation strategies, we will briefly go over the background information necessary to create high-performance software. In the following section, we will review the relevant aspects of the CPU hardware architecture that influenced \corrfunc's design.

\section{CPU Background: A Primer in CPU Architecture Design}\label{sec:cpu_primer}
\subsection{Evolution of CPU design towards multi-core processors}
Moore's law states that the number of transistors in an integrated circuit doubles roughly every two years \citep{Moore_1975}. This empirical observation has held up remarkably for well over 40 years from the 1970s to the mid-2010s. A corollary of Moore's law, as observed by Intel Corporation, is that CPU performance doubles every 18 months. This improvement in CPU performance comes from both faster and larger number of transistors on any given CPU. Computing throughput increases linearly with clock frequency, and therefore, all software benefited immediately from the higher clock frequencies. However, beginning in the early 2000s, CPU manufacturers began to run into issues with power dissipation. For a CPU with clock frequency $f$, the power required is $\propto f^3$, i.e., faster CPUs require a lot more power.
If this consumed power is not dissipated efficiently, then the temperature on the CPU would rise and cause the CPU to operate outside the thermal envelope --- a recipe for unstable CPUs and unreliable calculations. Cooling agents were not capable of dissipating the heat generated by the CPU quickly enough, and therefore the growth of clock frequencies stalled.
Hardware manufacturers had to seek out a different route to deliver higher computing throughput; the solution was CPUs with multiple cores. In the case of multiple cores, the power consumption only grows linearly with the number of cores. For example, two cores generate double computing throughput but only consume twice the power required by a single core.
A single core would need to run at twice the clock frequency and consume $4\times$ the power to provide the same computing throughput as two distinct cores. Naturally, hardware manufacturers then evolved towards multi-core CPUs, with each core operating at a lower clock frequency. This switch to multi-core CPUs represents a fundamental shift in the computing paradigm where increased computing capacity arises from a multitude of slower cores and not from faster individual cores.

\subsection{The CPU-Memory Performance Gap: The Emergence of the Cache Hierarchy}\label{section:caches}
In the initial stages of Moore's law, CPU clock speeds increased steadily. However, the increasing CPU clock speeds resulted in a hurdle -- to keep the CPU busy with computational work, the appropriate variables need to be available to the CPU. By design, CPUs only operate on data contained within a few hardware ``registers" located on the CPU. Since such CPU registers can only contain a {\em tiny} amount of data at any given time,\footnote{The latest generation Intel SkyLake CPUs contain 32 registers, each 512 bit wide.} a separate, larger memory storage is required to store the complete data during the computation. At a constant areal density of information in the substrate, increasing the memory on the CPU-chip would require a tremendous increase in the physical size of the CPU chip, with a correspondingly significant increase in the manufacturing costs. Therefore, the standard memory, or Random Access Memory (\RAM), had to be a physically distinct hardware component.  As a result, communication with the memory became slower,  especially relative to the continually increasing CPU speeds in the late 1990s to early 2000s.

Hardware designers tackled the ``slow-memory" issue by adding in cascading layers of faster memory closer to the CPU --- this is the so-called ``cache hierarchy". The smallest sized cache --- the level-1 (L1) cache\footnote{Here we will specifically focus on the $data$ cache, i.e., what {\em values} the CPU operates on. The L1i instruction cache contains what {\em operations} the CPU performs. Since the control flow of a typical program is usually linear, accessing the next instruction to execute is more predictable for the CPU.}, typically 64 KB --- was engineered physically closest to the CPU core and served as a dedicated cache for that core.  The next level, L2, is more substantial (typically in the range 4--20 MB) but slower.  The last-level cache (LLC), most frequently the L3 cache, has the largest size (20--40 MB) and is shared among all the physical cores on a single CPU.

Each cache adds a locality advantage over the next level down -- i.e., data found in the L1 cache is $2-5\times$ faster to retrieve compared to accessing from L2 and so on. A ``cache hit" occurs when data required for a compute operation are found in the cache, while a ``cache miss" implies data was {\em not} found in the cache. For any required data, the CPU will search through successive layers of cache, and if the data are still not found in the LLC, then a memory fetch is required to retrieve the data and populate the various cache levels. The cache hit-rate is {\em one of the most important} factors in determining how fast code will execute, with the largest speed-ups occurring when memory fetches can be replaced with L1 cache hits.

Let us consider the execution time of a hypothetical program that retrieves 100 values from memory. If each of these memory locations were {\em all} found in the L1 cache (i.e., 100\% cache hit-rate) and each L1 cache access takes 3 CPU cycles, then the program will complete in $100 \times 3 = 300$ CPU cycles. Now assume that the cache hit-rate has dropped to 95\%, and the data has to be fetched from L3 cache instead. If each L3 cache access takes 30 CPU cycles, the program will now take $95 \times 3 + 5 \times 30 = 435$ CPU cycles. A 5\% drop in the cache hit-rate resulted in a performance penalty of $135$ cycles, or $\sim 45\%$. Furthermore, consider the case where the cache misses required a fetch from \RAM instead of the L3 cache. Assuming a realistic value of 150 cycles for each memory read, the total program execution now becomes $95 \times 3 + 150 \times 5 = 1035$ CPU cycles. {\em A decrease of $5\%$ in the cache hit-rate results in a $\sim 3\times$ increase in the total runtime}. Thus, managing and increasing the cache hit-rate is a crucial feature for high-performance software.

\subsection{Singe Instruction Multiple Data -- \simd (Vectorisation)}
Once a code has implemented optimal cache management, performance can be improved further by parallelising the mathematical operations on arrays. Many computational tasks consist of element-wise operations on arrays of values, such as multiplying every element of an array by a scalar or adding two arrays element-by-element. Rather than operating on an element by element basis, such operations that are performed independently on each element can be executed by modern CPUs on blocks of elements, or vectors.  This paradigm is known as \simd --- Single Instruction Multiple Data\footnote{From a CPU design perspective, increasing performance by increasing the \simd width has compelling additional advantages. Wider \simd vectors only require a linear increase in power consumption~\citep{Rogozhin2017a} and avoid the cubic growth of CPU power consumption with frequency.}.
Standard \simd instruction sets include Streaming \simd Extensions (\sse) and Advanced Vector Extensions (\avx), and the more recent Intel Advanced Vector Extensions 512 (\avxft). The gains can be enormous - \avx code can operate on 8 \texttt{floats} simultaneously, and can potentially provide an $8\times$ speedup over un-vectorised code. The caveat is that to gain such speedups, the core computational logic has to be re-formulated by partitioning the computation in {\em independent} chunks of work.

\subsubsection{How to Create Vectorised Software}\label{sec:vectorized_code}
There are (at least) four different methods of creating vectorised software. In increasing order of implementation difficulty and typical performance improvement, these methods are:
\begin{enumerate}
\item {\em Writing for automatic vectorisation by compiler}: The essential requirement for automatic vectorisation is that each operation can be performed independently across the entire \simd width. That is, the final results do not depend upon the order of operations performed on a per-element basis within the \simd vector. Since the compiler has to generate correct code under {\em all} possible use-cases, the compiler is constrained to make conservative assumptions. Consequently, automatic vectorisation is strongly dependent on compiler, compiler version, compiler flags, as well as judicious use of \texttt{\#pragma} directives in the code.\footnote{\texttt{\#pragma} directives are used to mark code sections where vectorisation is safe and allows the compiler to relax the conservative assumptions.}
\item {\em Using \simd libraries}: There are some publicly available, general-purpose \simd libraries, e.g., \texttt{Vc}~\citep{vc_simd_lib_2011}, \texttt{vectorclass}(~\url{https://www.agner.org/optimize/#vectorclass}). These libraries wrap the different \simd math operations and abstract away the specifics of the underlying hardware instructions. Using these \simd libraries frequently require coding in \texttt{C++} but the resultant code resembles the familiar sequential code.
\item {\em Explicitly writing with \simd intrinsics}: Manipulate data directly in \simd data-types. Such codes may be portable but require intricate knowledge of instruction sets and various architectures.
\item {\em Writing in assembly language}: Use of assembly language produces the highest performance, but usually such implementations are not portable and additionally require detailed knowledge of hardware and compiler handling of inline assembly.
\end{enumerate}

The correlation function code contains a histogram update that constitutes a data dependency condition between consecutive iterations. Hence, correlation function codes can not be automatically vectorised. Implementing the \corrfunc code with \simd libraries (item ii) would have required additional knowledge of the \texttt{C++} language and would have delayed the creation of \corrfunc. Hence, we have used the \texttt{C} programming language
and explicit \simd intrinsics (item iii above) to create \corrfunc. We will discuss the \simd implementation in \S~\ref{sec:cell_simd}.

\section{Methods Overview}\label{sec:corrfunc_methods}
The \corrfunc framework is designed to tackle a broad swath of clustering problems. Given two sets of points, say source points\footnote{Since galaxies are treated as points within \corrfunc, we will use the terms ``points", ``particles" and ``galaxies" interchangeably} and query points, and a maximum separation, \corrfunc can quickly find the list of possible (query, source) pairs. While the \corrfunc framework was created to compute correlation functions efficiently, several other clustering statistics can benefit from such a framework. For example, the weak-lensing signal, $\Delta \Sigma$ (already implemented in \citealt{halotools}), counts-in-spheres, $pN(r)$, can be efficiently computed within the \corrfunc framework, as can pair-wise velocity dispersion~\citep{antonio_pvd}. Other algorithms that require a reduction over neighbours within a fixed separation, e.g., kernel density estimation, can also be efficiently implemented on top of the \corrfunc design. However, for the remainder of the paper, we will focus solely on the computation of correlation functions.

To summarise, a high-performance code must have these three characteristics within the compute-intensive sections:
\begin{enumerate}
\item predictable and contiguous memory access to benefit from the underlying hardware caches
\item vectorised operations to benefit from the wider vector registers present in modern CPUs
\item multi-core parallelism to benefit from the multiple cores in modern CPUs
\end{enumerate}
These three conditions only control how fast any given sequence of calculations are completed. In order to improve the absolute time to solution, any high-performance code should also minimise the total number of computations performed. For problems relevant to \corrfunc, the bulk of the computation time is spent on calculating pair-wise separations and then updating a histogram. The strategy, therefore, is to prune away as much of the potential search volume as possible and then look for further reductions at an individual particle level. We will describe how all of these optimisations are implemented within \corrfunc. In \S~\ref{sec:partition} and \S~\ref{sec:cell_refines} we will show how to reduce the absolute number of distance calculations by splitting the entire domain into multi-dimensional cells. In \S~\ref{sec:cellstruct} we show how contiguous memory access is ensured by first grouping all particles together in these multi-dimensional cells, and then calculating pair-wise particle separations within such cell-pairs. In \S~\ref{sec:cell_simd} we will show how the hand-written vectorised code works in \corrfunc. Finally, in \S~\ref{sec:openmp}, we will show how \openmp is implemented in \corrfunc.

\subsection{Reducing the Total Number of Distance Computations: Partitioning the particles on a grid}\label{sec:partition}
To compute a correlation function, we need to compute pair-wise separations and then count how many separations fall within some specified range $[\rmin, \rmax)$. Since the volume probed increases as the cubic power of the radius, the absolute number of particle pairs considered strongly depends on \rmax. For a given \rmax, to minimise the total number of distance computations, we would want to partition the domain into distinct spatial regions, and then quickly identify pairs of regions where particle-pairs {\em cannot} be within \rmax.
Most correlation function codes perform this spatial partitioning with either a tree structure or with a grid structure.

Tree structures map the entire domain into a ``root node", and then recursively sub-divide the domain into ``leaves" when some specified (maximum) threshold of particles is reached at any ``node". Such an adaptive, hierarchical spatial partitioning requires more complex tree construction algorithms and consequently, tree construction also frequently imposes a significant runtime overhead. Additionally, given an arbitrary query point, we may need to traverse several levels of the tree partitioning before locating the containing node (or leaf). Such a traversal amounts to accessing memory randomly and is detrimental to performance\footnote{Cache oblivious tree algorithms exist~\citep[e.g.,][]{vanEmdeBoas_tree_1975} but are even more complicated to design and implement. One open-source implementation can be found here: \url{https://github.com/lwu/veb-tree}.}.

In a grid structure, the entire domain is sub-divided into a grid with a fixed spatial width. Such a sub-division necessitates either knowing the entire domain extent a priori (e.g., for a cosmological box where positions must be within $0$ and the box-size) or, calculating the spatial extent by making a pass through the entire set of particles. Once the spatial extent is known, the partitions are simple axis-aligned sub-divisions of the specified width. Such a sub-division forms a crucial difference between the grid and the tree partitioning schemes. In the tree case, the number of partitions depends on both the spatial extent and the actual number and distribution of particles, while in the grid case the total number of partitions only depends on the domain extent and is independent of the number and distribution of particle positions. Because of the simplicity of the fixed width grid algorithm, the grid partitioning code is trivial {\em and} the runtime overhead for the grid partitioning is also lower than the tree case.

Compared to grids, tree structures like the
\textit{kd-tree}~\citep{Bentley75_kdtree} often have better theoretical scaling --- \onlogn or even \onpart. However, modern CPUs so strongly prefer the ordered memory access enabled by grid structures that often grids end up providing the faster time-to-solution.  In implementing \corrfunc, we initially tested tree algorithms and found that the simpler strategy of axis-aligned subdivisions outperformed tree algorithms.

Such cells have previously been used previously in astrophysics and named ``chaining-mesh" within the context of Particle-Particle-Particle-Mesh codes~\citep{chaining_mesh_p3m_hockney_73,p3m_eastwood80, hockney_and_eastwoord88,couchman_p3M_91}. Coarse grids have been used to demarcate potential regions of interactions in other fields of research as well --- just called differently. The coarse cells were called ``bin-lattice" while simulating the flocking behaviour of birds~\citep{Reynolds87_bin_lattice,Reynolds2000_bin_lattice}, ``cell linked-lists" or ``linked cell-list" in molecular dynamics~\citep{Quentrec_cell_linked_list73,Allen_linked_cell_list_orig_ref_89}.

In the following sections, we will discuss how to create two kinds of optimal space-partitioning grids -- one each for spatial and angular correlation functions. The spatial grid consists of axis-aligned partitions of a bounding cuboid, while the angular grid consists of partitions in latitude and longitude (declination and right ascension). The individual cell-sizes in both angular and spatial grids typically correspond to the maximum possible separations being probed.

\subsubsection{Partitioning the particles on a grid: Spatial correlation functions based on \rmax}\label{sec:grid}
We need to partition the particle domain so that we can efficiently prune the search volume that can not contain pairs within \rmax. As we discussed in \S~\ref{sec:partition}, we partition the domain using a 3D grid with a cell-width of (at least) \rmax. With such a grid, two points separated by more than one cell along any one dimension can not be within \rmax of each other.
The first step for creating the grid in Cartesian space is identifying the bounding cuboid. Since we might be performing cross-correlations involving two different data-sets, this cuboid should completely encompass both data-sets. For on-sky positions (i.e., in spherical coordinates), we first transform those positions into Cartesian coordinates and then compute the bounding box. Once we have the bounding box, the grid partitioning depends on the kind of separations required as well as the input catalogue type. For correlation functions defined on 3-D separations (e.g., \xiofr) for input catalogues defined in Cartesian space, we can set the cell-width along all axes to \rmax. For calculations requiring a line-of-sight separation on an input catalogue defined in Cartesian space (e.g., \wprp) we assume that the line-of-sight coincides with the $Z$-axis. Consequently, the cell-widths are \rmax in X and Y directions, and \pimax in the Z direction. For a similar calculation with on-sky positions, the line-of-sight will change for every galaxy pair considered, and hence, we can not assume that the line-of-sight to be axis-aligned (see the Appendix\ref{sec:appendix_separations} for the conventions for defining the pair-wise separations). Thus, for spatial correlation functions with on-sky positions, we set the cell-width along all axes to the maximum possible separation, $\mathcal{R}_{\rm sep, max} = \sqrt{\rmaxsqr + \pimaxsqr}$.

\begin{figure}
\centering
\includegraphics[clip=true, scale=0.9]{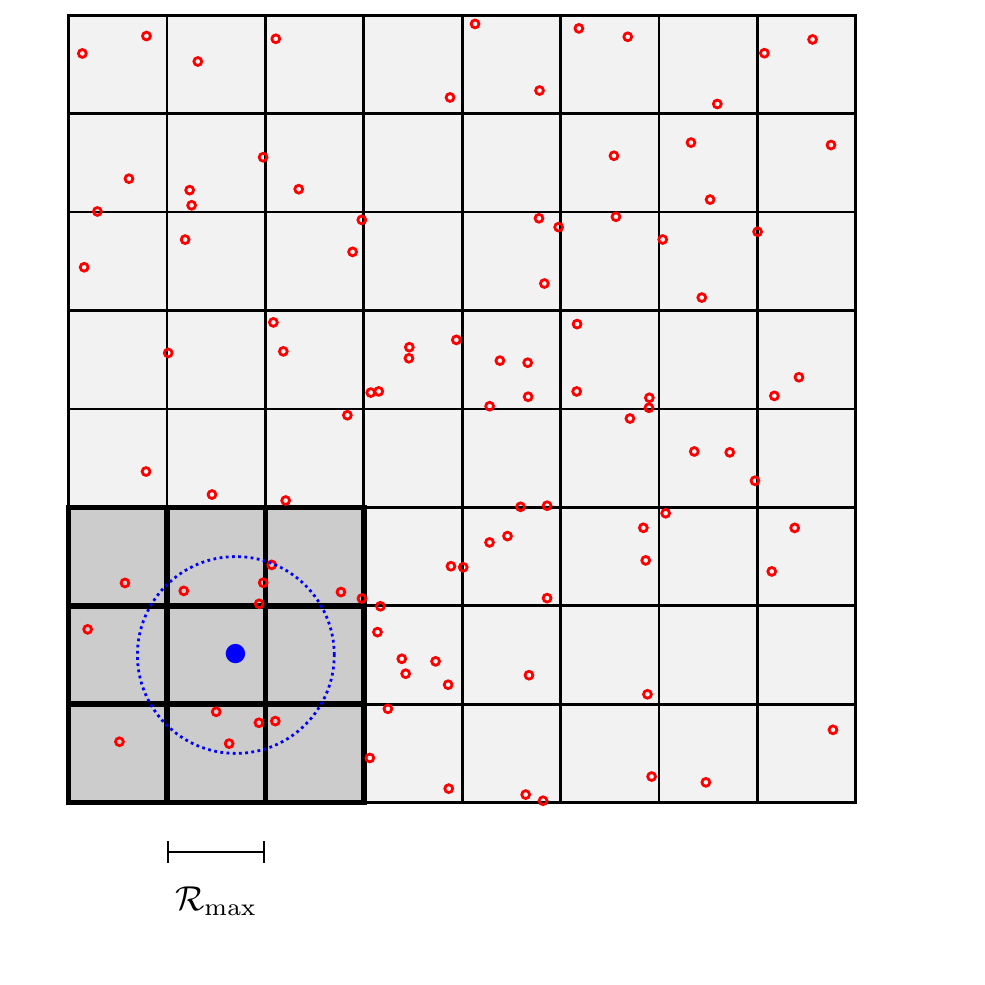}
\caption{A 2-D grid showing the bin-lattice partitioning scheme. The bigger square show the entire
domain, the red circles show a random distribution of 100 particles. Say we want to compute all pairs
for the target blue point, then we would only have to consider red points that are within one cell (the dark shaded region).
A circle with radius \rmax is also drawn to shown the actual pairs counted in the correlation function calculation.}
\label{fig:grid}
\end{figure}

Fig.~\ref{fig:grid} shows a schematic 2-D grid for the partitioning scheme. The
red circles represent the reference Poisson distributed points while the blue filled circle
shows the query point. Since the bounding cuboid and the total number of bins are the same for both datasets, any query point from a cell in the first dataset will fall into that same cell within the second dataset. Once the central cell is determined, {\em any} reference point that satisfies the distance inequality $\rsep < \rmax$, {\em must} lie
within the shaded nine cells in 2-D. Going to three dimensions, the total number of cells that need to be searched increases to 27.

Once we determine the maximum extent of both the particle distributions, we can compute the total number of cells along each dimension by dividing the extent by \rmax and taking the integral part. For example, if $(x_{max}, x_{min})$, represent the maximum and minimum along the $x$-axis, then the number of bins is: $n_x = \lfloor(x_{max} - x_{min}) \times \rmax^{-1}\rfloor$. Such a calculation ensures that the cell-width is {\em at least} \rmax in each dimension. The total number of bins is then $n_{tot} = n_x \times n_y \times n_z$. The $i$-th galaxy, with position $(x_i, y_i, z_i)$, will be located in the 3D cell specified by the following three indices:
\begin{align}
\begin{split}
i_x &= \left\lfloor(x_i - x_{min})/(x_{max} - x_{min}) \times n_x \right\rfloor\\
i_y &= \left\lfloor(y_i - y_{min})/(y_{max} - y_{min}) \times n_y \right\rfloor\\
i_z &= \left\lfloor(z_i - z_{min})/(z_{max} - z_{min}) \times n_z \right\rfloor\\
\end{split}
\end{align}

Thus, given the domain extent and a 3D $(n_x, n_y, n_z)$ grid, we can identify the precise cell that would contain any target galaxy. Since the cell width is at least \rmax by construction, any galaxy within \rmax of this target galaxy {\em must} be within the neighbouring cells. Therefore, we can immediately prune {\em all} of the cells (and the galaxies within those cells) not within one cell offset along each dimension~(see Fig.~\ref{fig:grid}).

\subsubsection{Partitioning the particles on a grid: Angular correlation functions based on \thetamax}\label{sec:sphere_grid}
Positions of observed galaxies are typically right ascension (\RA, $\alpha$) and declination (\DEC, $\delta$) and then possibly a line-of-sight distance. Where only the spatial correlation function is required, we can convert these spherical co-ordinates into equivalent Cartesian positions and perform the same lattice sub-division as shown in \S~\ref{sec:grid}. Since we need angular separations for an angular correlation function, we need a different method to partition the particles.
The angular separation between any two points can be computed by the Haversine formula:
\begin{align}
\begin{split}
\Delta \sigma = 2 \arcsin \sqrt{\sin^2\left(\dfrac{\Delta\delta}{2}\right) + \cos\delta_1\cdot\cos\delta_2\cdot\sin^2\left(\dfrac{\Delta\alpha}{2}\right)}\\
\implies \sin^2\left(\dfrac{\Delta\sigma}{2}\right) = \sin^2\left(\dfrac{\Delta\delta}{2}\right) + \cos\delta_1\cdot\cos\delta_2\cdot\sin^2\left(\dfrac{\Delta\alpha}{2}\right)
\label{eqn:ang_dist}
\end{split}
\end{align}
where, the subscripts $1, 2$ refer to the two galaxies and $\Delta \delta = |\delta_1 - \delta_2|$ and $\Delta \alpha = |\alpha_1 - \alpha_2|$, are the absolute differences in the \DEC and \RA respectively. For calculating an angular correlation function, we are only interested in point-pairs that lie within a certain \thetamax (typically $10\degree-20\degree$). However, in the general case, \thetamax has the range of $[0\degree, 180\degree]$ and the space-partitioning algorithm needs to be able to compute all possible angular separations. By design, \corrfunc enforces $\delta$ to be in the range $[-90\degree, 90\degree]$, implying that $\cos \delta >=0$ for all valid $\delta$. Then in Eqn.~\ref{eqn:ang_dist}, the second term in the right hand side is always $>=0$ for all allowed values of $\delta_1, \delta_2, \Delta\alpha$. Therefore, the maximum separation in \DEC that a pair can have while still being within \thetamax, occurs for $\Delta\sigma = \thetamax$ and can be written as:
\begin{align}
\begin{split}
\sin^2\left( \dfrac{\thetamax}{2}\right) &= \sin^2\left(\dfrac{\left(\Delta\delta\right)_\mathrm{max}}{2}\right),\\
\implies \left({\Delta \delta}\right)_\mathrm{max} &= \thetamax.
\end{split}
\end{align}
More physically, two particles that have have a \DEC separation of {\em higher} than \thetamax can {\em never} be within \thetamax. Immediately, we can see a binning strategy --- binning in $\delta$ with a bin-width of \thetamax and then looping over the neighbouring $\delta$ cells will be sufficient to capture all possible pairs within \thetamax.

We can further reduce the search volume by binning in \RA. We look at Eqn. ~\ref{eqn:ang_dist} again, and solve for maximum allowed value of $\Delta\alpha$ for any bin in $\delta$. The maximum value of $\Delta\alpha$ occurs when $\Delta\delta$ is 0 and $\cos\delta_1 \cdot \cos\delta_2$ has a minimum value allowed in the \DEC bin. Since $\cos\delta$ is monotonic in the allowed range of $\delta$, the minimum value occurs at the \DEC bin boundaries. This smallest value for $\cos\delta$ could be either the lower or upper edge of the \DEC bin; we take the smaller of the two. We denote this as $\left(\cos\delta\right)_\mathrm{min}$, and note that the minimum value is for the $\cos\delta$ and {\em not} $\delta$ itself. Setting $\Delta\delta=0$ and $\Delta \sigma = \thetamax$ in Eqn.~\ref{eqn:ang_dist}, we find
\begin{equation}
\begin{split}
&\phantom{\implies} &\sin^2\left(\dfrac{\thetamax}{2}\right) &=  \left(\cos\delta\right)_\mathrm{min}^2 \cdot \sin^2\left(\dfrac{(\Delta\alpha)_\mathrm{max}}{2}\right),\\
&\implies &\sin\left(\dfrac{\left(\Delta\alpha\right)_\mathrm{max}}{2}\right) &= \dfrac{\sin\left(\dfrac{\thetamax}{2}\right)}{\left(\cos{\delta}\right)_\mathrm{min}}, \\
&\implies &\left(\Delta\alpha\right)_\mathrm{max} &= 2 \cdot \arcsin \left( \dfrac{\sin\left(\dfrac{\thetamax}{2}\right)}{\left(\cos{\delta}\right)_\mathrm{min}} \right).
\end{split}
\label{eqn:ra_binning}
\end{equation}
For cases where $\left(\cos\delta\right)_\mathrm{min}$ is close to 0 (i.e., close to the poles), we fix $(\Delta\alpha)_\mathrm{max}$ to be the entire \RA range of the particle distribution, i.e., only one \RA grid cell is constructed in such a \DEC bin. While the \RA binning is done by default in \corrfunc, there is a runtime option to disable the \RA binning. In that case, the angular partitioning only occurs in \DEC.

\begin{figure}
\includegraphics[width=0.65\linewidth,center]{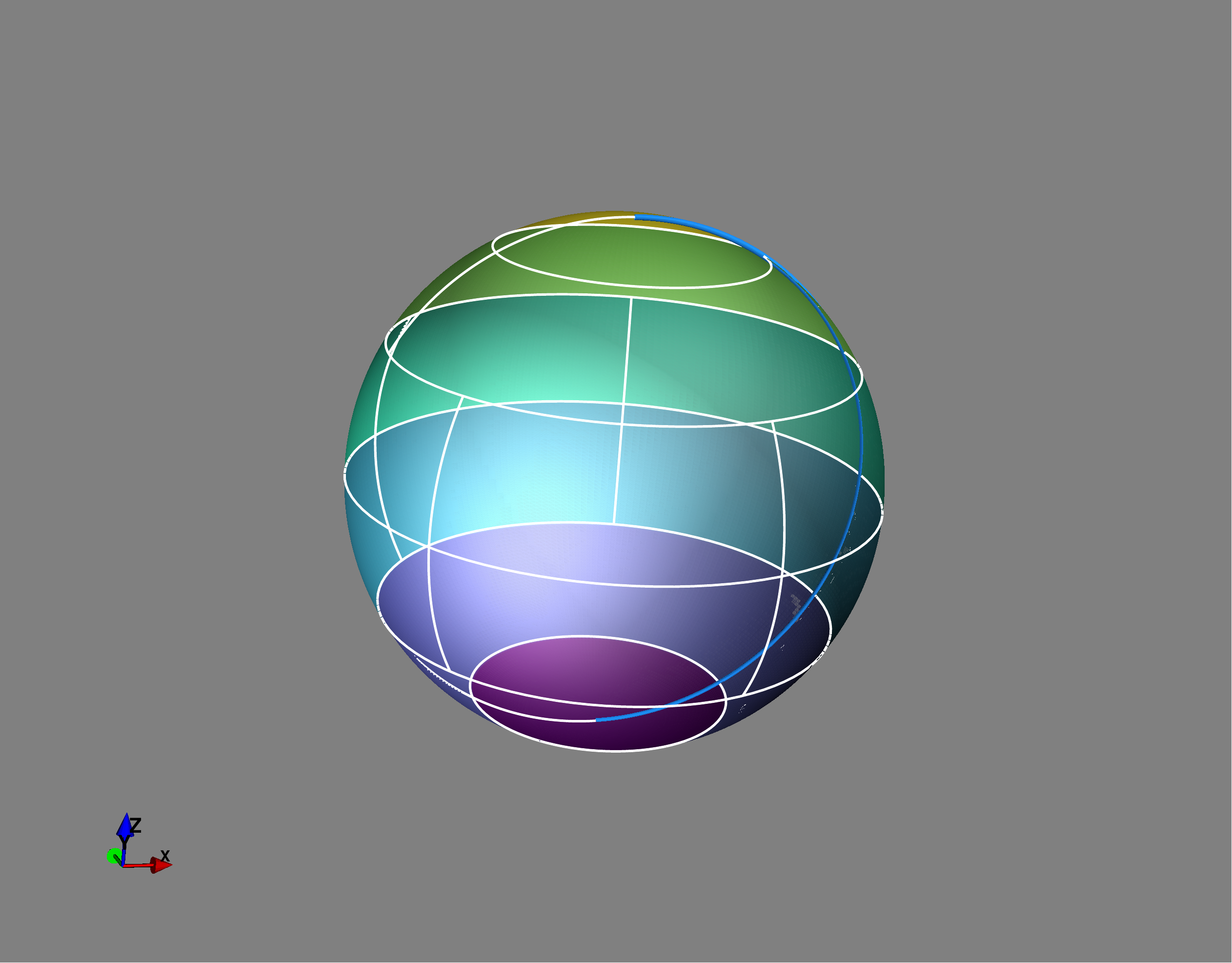}
\caption{\small Spherical grid for \thetamax=30\degree and linking in \RA enabled. The blue longitude represents the minimum and maximum of the \RA limits; in this example, the minimum and the maximum \RA limits coincide but are not required to be the same. The script for creating this interactive plot is here: \protect\url{https://github.com/manodeep/Corrfunc/blob/develop/paper/figures/generate_sphere_grid.py}.}
\label{fig:sphere_grid}
\end{figure}

\subsection{Reducing the Total Number of Distance Computations: Sub-dividing the Cells}\label{sec:cell_refines}
In the previous section, we set the cell-widths along each axis to be the maximum separation along that axis. In this section, we will show how to optimise that partitioning scheme further, specifically focusing on the spatial correlation functions (similar arguments also apply for \wtheta).
With the spatial partitioning methods described above, we have approximated the spherical search volume of $\dfrac{4}{3}\pi \rmaxcubed$ with $\left(3\rmax\right)^3$. For 2-D correlation functions, the cylindrical search volumes of $\pi\rmaxsqr\times2\pimax$ is replaced with $\left(3\rmax\right)^2\times(3\pimax)$
and $\left(3\sqrt{\rmaxsqr + \pimaxsqr}\right)^3$ for input positions in Cartesian and spherical coordinates
respectively. For randomly distributed particle positions, the ratio of the actual search volumes to the minimum search volume directly probes the fraction of spurious pair-wise separations. If we keep the cell-width exactly \rmax,
then only $\dfrac{4}{3}\pi\rmaxcubed/(27\rmaxcubed) \approx 16\%$ of the total number of pair-wise separations will satisfy the distance constraint. However, since we can only have an integral number of cells, the cell-width will be greater than \rmax -- further increasing the search volume. Therefore, in practice, {\em at least} 84\% of the total separations computed will have to be discarded. For 2-D separations with input Cartesian positions, only $2\pi/27 \approx 24\%$ of the separations will be within the requested distance cuts.

For input positions in spherical coordinates, the situation can be even worse. Assuming $\pimax \sim \beta\times\rmax$, only $2\beta\pi/\left(27 \times \sqrt{1+\beta^2}\right) \approx 2\pi/27 \times \left(1 - \dfrac{1}{2\beta^2}\right)$. For realistic correlation functions, $\beta$ will be in the range $2-4$, and correspondingly, the fraction of valid separations will be $20-24\%$. Thus, for 2-D positions, $\sim 75-80\%$  of all the distance computations will have to be discarded if the default cell-widths are {\em at least} the maximum separation requested.

\begin{figure}
\centering
\begin{tikzpicture}[]
\usetikzlibrary{patterns}
\def\multfac{0.35}
\pgfmathsetmacro\rectx{\multfac*\linewidth}
\pgfmathsetmacro\recty{\rectx}
\pgfmathsetmacro\halfrectx{\rectx*0.5}
\pgfmathsetmacro\thirdrectx{\rectx/3.0}
\pgfmathsetmacro\leftx{-\rectx}
\pgfmathsetmacro\xstart{\leftx}
\pgfmathsetmacro\ystart{0}
\pgfmathsetmacro\arrowshorten{4 pt}
\pgfmathsetmacro\ystagger{0.15*\linewidth}
\pgfmathsetmacro\yoff{-\ystagger - \recty}

\coordinate (A) at (\xstart,{\ystart pt});
\draw[ultra thick] (A) rectangle ++({\rectx pt},{\recty pt}) coordinate (B);
\draw[ultra thick] (B) rectangle ++({\rectx pt},{-\recty pt}) coordinate (C);

\path[<->, >=latex,thick, shorten >=\arrowshorten,shorten <=\arrowshorten] (A) edge[bend right=20] node[below] {$\rmax$} ([xshift={\rectx pt}]A.west);
\path[<->, >=latex,thick, shorten >=\arrowshorten,shorten <=\arrowshorten] ([xshift={\rectx pt}]A.west) edge[bend right=20] node[below] {$\rmax$} (C);

\path (A) ++(0,{\yoff pt}) coordinate (AA);
\draw[ultra thick] (AA) rectangle ++({\rectx pt},{\recty pt}) coordinate (BB);
\draw[dashed, ultra thick] ([xshift={\halfrectx pt}]AA.east) -- ([xshift={-\halfrectx pt}]BB.west);
\draw[ultra thick] (BB) rectangle ++({\rectx pt},{-\recty pt}) coordinate (CC);
\draw[dashed, ultra thick] ([xshift={\halfrectx pt}]BB.east) -- ([xshift={-\halfrectx pt}]CC.west);

\path[<->, >=latex,thick, shorten >=\arrowshorten,shorten <=\arrowshorten] ([xshift={\halfrectx pt}]AA.east) edge[bend right=20] node[below] {$\rmax$} ([xshift={-\halfrectx pt}]CC.west);

\path (AA) ++(0,{\yoff pt}) coordinate (AAA);
\draw[ultra thick] (AAA) rectangle ++({\rectx pt},{\recty pt}) coordinate (BBB);
\draw[dashed, ultra thick] ([xshift={\thirdrectx pt}]AAA.east) -- ([xshift={-2.0*\thirdrectx pt}]BBB.west);
\draw[dashed, ultra thick] ([xshift={2.0*\thirdrectx pt}]AAA.east) -- ([xshift={-\thirdrectx pt}]BBB.west);
\draw[pattern=north west lines, pattern color=gray, thin, draw=none] (AA) rectangle ++({\halfrectx pt},{\recty pt});
\draw[pattern=north west lines, pattern color=gray, thin, draw=none] (CC) rectangle ++({-\halfrectx pt},{\recty pt});

\draw[ultra thick] (BBB) rectangle ++({\rectx pt},{-\recty pt}) coordinate (CCC);
\draw[dashed, ultra thick] ([xshift={\thirdrectx pt}]BBB.east) -- ([xshift={-2.0*\thirdrectx pt}]CCC.west);
\draw[dashed, ultra thick] ([xshift={2.0*\thirdrectx pt}]BBB.east) -- ([xshift={-\thirdrectx pt}]CCC.west);

\draw[pattern=north west lines, pattern color=gray, thin, draw=none] (AAA) rectangle ++({2.0*\thirdrectx pt},{\recty pt});
\draw[pattern=north west lines, pattern color=gray, thin, draw=none] (CCC) rectangle ++({-2.0*\thirdrectx pt},{\recty pt});

\draw[pattern=north east lines, pattern color=gray, thin, draw=none] ([xshift={\thirdrectx pt}]AAA.east) rectangle ++({\thirdrectx pt},{\recty pt});
\draw[pattern=north east lines, pattern color=gray, thin, draw=none] (CCC) rectangle ++({-\thirdrectx pt},{\recty pt});

\path[<->, >=latex,thick, shorten >=\arrowshorten,shorten <=\arrowshorten] ([xshift={\thirdrectx pt}]AAA.east) edge[bend right=20] node[below] {$\rmax$} ([xshift={\thirdrectx + \rectx pt}]AAA.east);
\path[<->, >=latex,thick, shorten >=\arrowshorten,shorten <=\arrowshorten] ([xshift={2.0*\thirdrectx pt}]AAA.east) edge[bend right=20] node[below] {$\rmax$} ([xshift={-\thirdrectx pt}]CCC.west);
\end{tikzpicture}

\caption{\small A schematic demonstrating how refining the grid reduces the search volume. In top panel, all possible pairs between the left and right sections will have to examined for potential pairs. In the middle panel, the cells are sub-divided into two (i.e., cell-width of \rmax/2) and the particles in the left-most and right-most sections can never be within \rmax. In the bottom panel, the cells are further sub-divided into three (i.e., cell-width of \rmax/3). As the arrows marking \rmax at the bottom of the panel show, The two right-most sections can not have a particle pair with the left-most section, and the right-most section can not have a pair with the two left-most sections.}
\label{fig:grid_refine}
\end{figure}
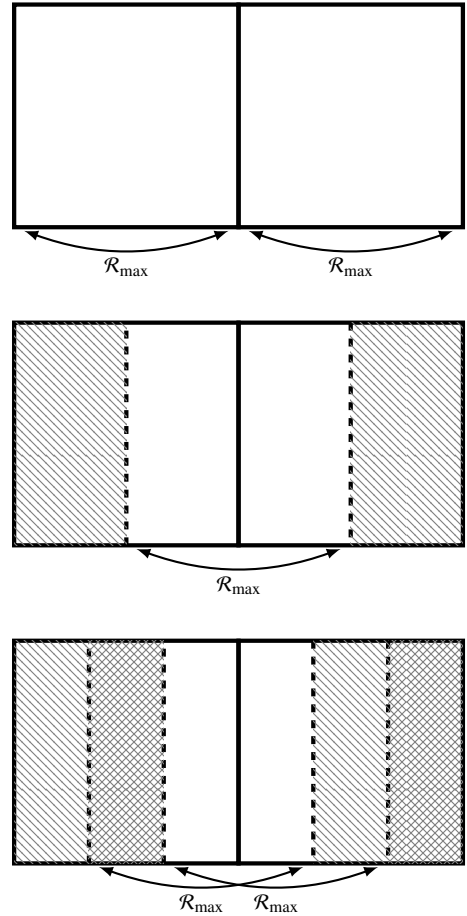

One straightforward method to reduce the number of un-needed calculations would be to refine the grids further. Following \citet{cell_refinements_gonnet_07}, we show how to reduce the search volume by sub-dividing the cells. In the first panel of Fig.~\ref{fig:grid_refine}, we use \rmax as the default size.  Consequently, separations have to be computed for all possible particle pairs between the two cells. If $n_\mathrm{cell}$ is the average number of particles per cubical cell of side \rmax, then the total number of distance computations will be $n_\mathrm{cell}^2$. In the middle panel, we reduce the cell-width to \rmax/2; now we can see that the two outer-most cells (hatched regions) cannot be within \rmax. Similarly, in the bottom panel, we show the case for a cell-width of \rmax/3. Comparing the middle and bottom panel, we can see that if we keep sub-dividing the cells, we can keep reducing the search volume. As shown in \citet{cell_refinements_gonnet_07}, sub-dividing the \rmax cells into $k$ sub-cells, reduces the total computations to $(k + 1)/2k \times n_\mathrm{cell}^2$, where $n_\mathrm{cell}$ is the average number of particles per cell. In the limit ${k \to \infty}$, the total number of computation becomes $1/2 \times n_\mathrm{cell}^2$, a factor of 2 improvement over the case with cell-width of \rmax.\footnote{This limiting case is similar to a tree structure where each leaf has at most one particle.} However, if we sub-divide a \rmax cell into $k$ sub-cells, then we have to loop over a set of $(2k + 1)$ neighbour cells along that dimension. Assuming that each cell is now sub-divided into $(k_x, k_y, k_z)$ sub-cells, then the total number of neighbouring cells that have to be checked are $(2k_x + 1) \times (2k_y + 1) \times (2k_z + 1)$. For the case of $(k_x, k_y, k_z) = (1, 1, 1)$, we have to inspect 27 neighbouring cells, while for $(k_x, k_y, k_z) = (2, 2, 2)$, we have to inspect 125 neighbouring cells. As we increase the number of sub-cells, $k$ per \rmax cell, the overhead of looping through the neighbouring cells increases rapidly. The trade-off is in finding a suitable $k > 1$, that balances between the two asymptotic runtime behaviour cases and produces the fastest code.

\subsection{Reducing the Total Number of Distance Computations: Associating Pairs of Cells}\label{sec:cell_pairs}
Once all the particles are assigned to cells, we need to associate pairs of cells that {\em may} contain particles within the maximum separation. For any given cell, say the ``source" cell, we need to identify all possible ``target" cells that can contain particles within the maximum separation.

For spatial grids, the individual cell-widths along each dimension are constrained to be at least the maximum possible separation (except when refining the grid; see the previous section). Therefore, all possible particle pairs can be separated by {\em at most} one cell along each Cartesian axes.

When periodic boundary conditions are requested, we wrap the target cell indices into $[0, n_x)$. For example, for a source cell with X-index $0$, the cell to the immediate left along the X-axis, i.e., a target cell index of $-1$ will be mapped to the target cell with index $(n_x-1)$. This wrapping is performed along all three axes, and whenever such wrapping is required, we also store the offset required to move the source cells adjacent to the target cell. In the example above, we would need a spatial offset of \lbox\footnote{For a periodic wrap going the other direction we store a spatial offset of -\lbox. We will refer to these offsets in \S\ref{sec:pbc} as $\Delta_\mathrm{X}$.} along the X-axis to linearly translate the source cell closer to the target cell. When periodic wrapping is not required for a cell-pair, including calculations without periodic boundary conditions, this spatial offset is identically set to 0. We will need this spatial offset for an optimised implementation for calculating separations under periodic boundary conditions (see \S~\ref{sec:pbc}). We store a total of three spatial offsets -- one per axis -- for each cell-pair.

For angular correlation functions when only binning on \DEC behave similarly to the spatial correlation function. We can further sub-divide each \DEC bin and benefit from the reduced search volumes. The situation is little more tricky when \RA gridding is enabled since the \RA cell-widths depend on the \DEC band (see in Eqn.~\ref{eqn:ra_binning}). Therefore, \RA cell-widths on distinct \DEC bands are likely to be different. By construction, the first \RA cell for all \DEC bands are constrained to begin at $0\degree$. Therefore, the \RA grids do not necessarily form continuous edges across \DEC bands (see Figure~\ref{fig:sphere_grid}). To account for such staggered \RA binning, we need to consider two extra cell-pairs --- one in the clockwise and another in the counter-clockwise rotation --- when associating cell-pairs across different \DEC bands. Effectively, we are projecting the east and the west \RA boundaries from the source cell on to the target \DEC band. These new projected cells now identify the minimum and maximum extent for the source cell on the target \DEC band\footnote{We carry the actual spatial extents derived from the particle positions in that source \RA cell rather than the fixed grid extents.}. We can then identify the number of \RA cells on either side of these projected \RA cells based on the \RA cell-width at the target \DEC band. Since the \RA cells are periodic, and the same target cell might be considered multiple times (depending on the binning strategy), we only retain unique target cells for any given source cell.

\subsection{Reducing the Total Number of Distance Computations: Symmetry considerations for Autocorrelations}\label{sec:cell_autocorr}
When computing auto-correlations, we can can optimise further and reduce the total number of cell-pairs. Since we only have a single data-set for an auto-correlation, we can compute all particle pairs by first only computing unique particle-pairs and then doubling the resultant pair-counts. To compute unique particle pairs, we need to consider unique cell-pairs, as well as unique particle pairs within the same cell. When associating pairs of cells (see  \S~\ref{sec:cell_pairs}), we only associate unique cell pairs for auto-correlations. We ensure unique cell pairs by only keeping cell pairs where the array index for the target cell exceeds the array index for the source cell, i.e., prune up to half of all possible neighbour cell-pairs. To count only the unique particle pairs within the same cell, we consider target particle indices that are larger than the source particle index. Once all calculations are complete, we double the total pair-count for each bin and arrive at the all possible pair-counts. This strategy ensures that the auto-correlation of a data-set yields the same result as a cross-correlation with itself.  There is a small caveat to this doubling --- for cases where the smallest requested separation is $0.0$ we need to include self-pairs of particles. However, this self-pair for a particle should {\em not} be counted twice; therefore, we explicitly correct the pair-counts for the $\rmin == 0.0$ case. We show the justification for the zero minimum separation case in Appendix~\ref{appendix:self-pairs}.

\subsection{Speeding up the Calculation of Separations in Cell-Pairs: Improving Cache Locality}\label{sec:cellstruct}
In \S~\ref{sec:partition}, we discussed how to partition the particles into spatial and angular cells, and in \S~\ref{sec:cell_pairs} we showed how to associate only cell-pairs that may contain a particle pair. Given such a cell-pair, we still have to potentially inspect all possible particle pairs formed between particles in these two cells. Therefore, we need to quickly perform two tasks --  (i) identify the containing cell for any given particle and (ii) identify all particles belonging to any cell. Extracting optimal performance from the modern CPU architecture (see \S~\ref{section:caches} and \ref{sec:vectorized_code}) necessitates that the particles within a cell are stored contiguously. The input order of galaxies is likely to be arbitrary and unrelated to the correlation function calculation. Thus, we need to access the galaxies in a manner different from the input sequence.
For any given cell, we could store an array containing the original indices (i.e., input order) of all the particles belonging to that cell. Since we only have to store one number per particle -- the input array index -- such a strategy would increase the memory footprint only linearly with the total number of particles. However, since the particles are likely in arbitrary order, if we only contiguously store the particle {\em indices} within each cell, the particle positions that cell would still be located at vastly different memory addresses. The resultant memory access of the linked cell-list resembles random memory access, and we lose all benefits from the caching mechanisms present in modern CPUs.
Except for re-ordering the input arrays~\footnote{We have implemented an option to re-order the input particle arrays based on this ``cell-index" array in \corrfunc \texttt{v2.3.0}.}, the only way to achieve a contiguous memory layout for galaxy positions is by creating a duplicate of the particle data. We ensure particles are sequential in memory by duplicating the all the particle arrays and storing particles in the same cell within dedicated, contiguous arrays. We ensure that the particle locations are contiguous by moving them into the \texttt{C struct} (see Code~\ref{code:c-struct}) in the input order.
\begin{listing}
\caption{\small{The definition of the structure that holds all the points within {\em each} cell and ensures contiguous memory access. The variable type \texttt{DOUBLE} gets processed both as \texttt{float} and \texttt{double}, and the appropriate structure is selected at runtime.}}
\begin{minted}[escapeinside=||]{C}
struct cellarray_DOUBLE{
  |\textcolor{vargreen}{\textbf{DOUBLE}}| *x;
  |\textcolor{vargreen}{\textbf{DOUBLE}}| *y;
  |\textcolor{vargreen}{\textbf{DOUBLE}}| *z;
  |\textcolor{vargreen}{\textbf{int64\_t}}| nelements;
};
\end{minted}
\label{code:c-struct}
\end{listing}
After all the particles are assigned to cells, processing any cell-pair uses these \texttt{x/y/z} arrays associated with each cell.
Since the typical particle data is small ($\sim$20--50 MB), duplicating the entire particle list within the cells does not pose a strong memory requirement. However, accessing sequential memory locations provides a significant performance boost in modern CPUs, and justifies the added memory overhead.

\subsection{Speeding up the Calculation of Separations in Cell-Pairs: Accounting for Periodic Boundary conditions}\label{sec:pbc}
Periodic boundary conditions are almost always used in cosmological simulations of structure formation. For a cosmological box of size \lbox, particle positions are always constrained to lie in the closed interval $[0.0, \lbox]$. Periodic wrapping means that the maximum possible separation along any one particular axis is constrained to be $\leq 0.5\lbox$, and correspondingly, all numerically larger separations need to be correctly wrapped.
\begin{listing}
\caption{\small{C code to show how periodic boundary conditions are usually implemented in the context of calculating pair-wise separations.}}
\begin{minted}[escapeinside=||]{C}
for(int64_t i=0;i<N1;i++) {
    for(int64_t j=0;j<N2;j++) {
        |\textcolor{vargreen}{\textbf{DOUBLE}}| dx = |$x_i - x_j$|;
        if (dx > 0.5*|\lbox|) dx -= |\lbox|;
        if (dx <= -0.5*|\lbox|) dx += |\lbox|;
    }
}
\end{minted}
\label{code:normal_pbc}
\end{listing}
In the usual implementation of periodic boundaries (see Code~\ref{code:normal_pbc}), first the separation is computed between the pair of points and then the separation is compared against $0.5\times\lbox$, and finally the separation is wrapped where necessary. This approach requires at least two comparisons {\em per} axis, and then requires one additional addition to implement the wrapping (see Code~\ref{code:normal_pbc}). All of these calculations would have to be performed for every separation computed. However, within \corrfunc we know a priori whether or not periodic boundary conditions will be applicable when associating pairs of cells (see \S~\ref{sec:cell_pairs}). When associating cell-pairs, we additionally store a spatial offset along each axis. With this offset (say $\Delta_\mathrm{X}$), we linearly translate the coordinates for the particles in the first cell closer to the second cell (see Code~\ref{code:corrfunc_pbc}).
\begin{listing}
\caption{\small{C code to show how to avoid explicit periodic wrapping in the inner loop by transforming the positions of the particles in the first cell.  For cell-pairs that require periodic wrapping, $\Delta_\mathrm{X}$ is set to $\pm \lbox$, otherwise  $\Delta_\mathrm{X}$} is identically set to 0.}
\begin{minted}[escapeinside=||]{C}
for(int64_t i=0;i<N1;i++) {
    |\textcolor{vargreen}{\textbf{DOUBLE}} $x^{\prime}_i = x_i$ + $\Delta_{\mathrm{x}}$|;
    for(int64_t j=0;j<N2;j++) {
        |\textcolor{vargreen}{\textbf{DOUBLE}}| dx = |$x^{\prime}_i - x_j$|;
    }
}
\end{minted}
\label{code:corrfunc_pbc}
\end{listing}
With this formulation, we can avoid both the (computationally expensive) \texttt{if} conditions, and the consequent floating point addition for periodic wrapping for {\em each} pair. Since the translated position for the $i$-th particle only needs to be calculated {\em once} for {\em all} the \texttt{N2} particles, the overhead imposed by this extra addition is minimal.

\subsection{Speeding up the Calculation of Separations in Cell-Pairs: Sorting to enable Late Entry and Early Exit Conditions}\label{sec:cell_sorting}
The final step to reduce the total number of computations {\em per} cell-pair involves finding opportunities for early terminations or avoiding distance to bin calculations altogether.
To reduce the total number of computations involving particles in the second cell, we need to prune all particle pairs that cannot be within \rmax. For instance, if we knew the initial set of particles in the second list that cannot be within \rmax, then we could avoid retrieving the positions (and potentially weights) associated with that entire sub-set. Similarly, we would like to identify if no further pairs are possible for all remaining particles in the second data-set\footnote{We extend this early exit to the first data-set as well in \cite{corrfunc_exascale_2018} and \corrfunc \texttt{v2.3.0}.}. We implement both these optimisations by sorting the particles based on their $z$ positions.

To compute all possible particle pairs between two cells, we need a nested double \texttt{for} loop. We will adopt the notation that $i$-loop refers to the particles in the first cell, while the $j$-loop refers to particles in the second cell. With this convention in mind, consider the case of a pair of particles. Let $dz:=z_j - z_i$ be the separation between an $i$ and $j$ particle. From the triangle inequality, the total 3-D separation must be at least $dz$.
Therefore, particles with $\left | dz \right | \ge \rmax$ must have total separation of at least \rmax.\footnote{We can equivalently replace \rmax with \pimax for 2D correlation functions}. Thus, we only need to consider particle pairs that satisfy: $-\rmax < dz < \rmax$. Since the particles are sorted in increasing $z$, then the difference  can not decrease for constant $i$ (i.e., constant $z_i$) and increasing $j$ (i.e., increasing $z_j$). Therefore, we can only have a potential particle pair with the $i$-th particle in the first cell with the $j$-th particle in the second cell when $dz > -\rmax$. Similarly, when $dz \ge \rmax$ {\em no} further $j$-particle can have a pair with the current $i$-th particle. Taken together, these two conditions provides the basis for a late entry to and an early exit from the $j$-loop. Algorithmically, for a given $i$-particle in the first cell, we access only the $z_j$ positions and increment $j$ until $dz > -\rmax$. Once this inequality is satisfied, we access the full spatial positions of both particles to compute the relevant separation, and continue with the $j$-loop until we encounter $dz \ge \rmax$.
We reduce the search volume from $3\pimax$ to $2\pimax$ with such a sorting of particles in the $z$-direction (see \S~\ref{sec:cell_refines} for the discussion on the search volumes).

\subsection{Speeding up the Calculation of Separations in Cell-Pairs: Explicit Vectorisation}\label{sec:cell_simd}
So far, we have focused on two optimisation themes -- reducing the total search volume and improving the memory access pattern.
We have discussed how to reduce the total search volume -- by partitioning the particles into cells (\S~\ref{sec:partition}), refining the cell-widths (\S~\ref{sec:cell_refines}), only associating pairs of cells that may  have a pair within $\rmax$ (\S~\ref{sec:cell_pairs}), and then further reducing the total number of cell-pairs using symmetry considerations when computing auto-correlations (see \S~\ref{sec:cell_autocorr}). Once we do have a pair of cells, we showed how to reduce the memory access times by ensuring contiguous particle layout (\S~\ref{sec:cellstruct}), and then minimising memory traffic by implementing a `late entry' and `early exit' conditions (\S~\ref{sec:cell_sorting}). Now the remaining optimisation is writing explicitly vectorised code for computing the pair-wise separations between the $i$-th particle and all possible pairs with the remaining particles in the second data-set.

As we described in \S~\ref{sec:vectorized_code}, an essential condition for automatic vectorisation is that the final result should not depend on the the order in which each element of the vector register is processed. The mandatory histogram update (see Code~\ref{code:naivecorr}) in a correlation function breaks this requirement of unordered operations -- updating any particular histogram bin can only occur {\em after} all previous updates to the {\em same} bin have completed. If we perform the histogram update in parallel, then the output histogram would depend on the individual values contained in the \simd vector, and therefore, can not be guaranteed to be consistent (at compile-time) with sequential processing. Thus the compiler can not generate vectorised code for any calculation involving a histogram update.
Consequently, the tested compilers (\texttt{gcc}, \texttt{clang}, \texttt{icc})
could not automatically vectorise (see \S~\ref{sec:vectorized_code}) the default correlation function code.\footnote{In the \corrfunc kernels, we find that even the latest \texttt{AVX-512CD} instruction set (``Conflict Detection" subset of \avxft), designed specifically to vectorise histogram updates, does not result in vectorised code for the \corrfunc kernels.}. The various \texttt{C++} vector libraries proved too difficult to adapt without a sufficient knowledge of \texttt{C++}. Hence we had to resort to explicitly programming with vector intrinsics.

\simd vector intrinsics process \simdlen chunks of elements at once. As we described in \S~\ref{sec:vectorized_code}, the size of the vector registers are fixed -- 256 bytes and 128 bytes for \avx and \sse respectively. Therefore, the number of elements processed at a time, i.e., \simdlen, depends on the size of an individual element. For elements of type \texttt{float32}, 8 (4) items can fit into the 256 (128) bytes corresponding to the \avx (\sse) vector register. Similarly, 4 (2) elements of type \texttt{float64} can be processed simultaneously with \avx (\sse) instructions.

Computing a correlation function requires a double-nested \texttt{for} loop (see Code~\ref{code:naivecorr}). For vectorising this calculation, we need to re-write the algorithm with explicit vector intrinsics covering the computations in the second \texttt{for} (i.e., the $j$) loop. Note that the initial $j$-value at the beginning of the loop corresponds to the first $j$-particle that satisfies $dz \ge -\rmax$ (see \S~\ref{sec:cell_sorting}). The vectorisation is relatively straight-forward and consists of loading \simdlen chunks of $j$-particles, and computing the pair-wise separations with a fixed $i$-particle. We can then locate and update the appropriate histogram bin for each of the \simdlen pair separations. Any $j$-particles left after the \simdlen chunks are then processed in a scalar remainder loop\footnote{The biggest challenge in creating the \simd kernels was in figuring out the correct syntax for issuing the vector intrinsics and we consulted the Intel Intrinsics Guide extensively~(\url{https://software.intel.com/sites/landingpage/IntrinsicsGuide/}).}.

One of the novelties in the vectorised kernels within \corrfunc implementation is that a single source file works for both \texttt{float32} and \texttt{float64} arrays. Adding such flexibility (equivalent to \texttt{C++ templates}) is easier for sequential code, the \simd kernels pose more of a challenge from the conversion of the \simd registers into boolean masks~(see \S~\ref{sec:cell_popcnt}) and then further into \texttt{C integers}. We achieve this dual-precision functionality by heavily using \texttt{C macros}. For each kind of supported \simd operation (say, addition, multiplication), we have created a custom ``macro" and all the \simd operations within the \corrfunc vector kernels are written using these custom \simd macros. At compile-time, each source file is pre-processed to generate two different sources -- one for \texttt{float32} and another for \texttt{float64} input arrays. The custom \simd macro then expands to the appropriate \simd instruction at compile-time, resulting in two dedicated functions that each target one of the \texttt{float32} and \texttt{float64} operations. This flexibility is abstracted away from the user, and the \corrfunc interface detects the input array-type and offloads the calculation to the appropriate function. We are hopeful that the macros we have created (in the files \texttt{utils/avx\_calls.h} and \texttt{utils/sse\_calls.h} and \texttt{utils/function\_precision.h}) will enable other researchers to write their own \simd kernels for compute-intensive codes.

\subsection{Speeding up the Calculation of Separations in Cell-Pairs: Finding and Updating the Histogram Bins}\label{sec:cell_popcnt}
Once we have a pair separation that satisfies both the minimum and maximum separation cuts, we have to locate and increment the appropriate histogram bin.  \corrfunc only requires that the histogram bins be monotonically increasing and contiguous --- i.e., where the lower bound of a bin is the same as the upper bound of the previous bin. Because we allow arbitrary bin-widths, there is no direct way to calculate the histogram bin index for a given separation. The easiest way to locate the bin index for a given separation is to loop through the bins until the separation falls within the bin edges. To avoid the expensive \texttt{sqrt} operation for every pair, we locate the histogram bin index with squared distances. Similarly, for \ddtheta, we replace the angular bins in $\theta$ with equivalent bins in $\cos\theta$. With such a replacement we can avoid the computationally expensive \texttt{arccos} operation within the inner loop and significantly speed up the code.
\footnote{If the user does not request the average separation, then the \texttt{sqrt} and
\texttt{arccos} operations are skipped entirely.}

We can optimise the histogram bin lookup further based on the underlying physical scenario for correlation functions. Since the bins are sorted in increasing order, later bins encompass larger volumes and therefore should contain a larger number of pairs. Therefore, a priori we expect that separations are much more likely to fall into the final bins rather than the initial ones. Thus, we loop through the histogram bins in reverse, update the histogram bin count if necessary and break from the histogram update loop when there are no further valid particles left. We implement this backwards looping strategy for both the \simd and scalar sections. This optimisation reduces the total number of times the histogram loop is executed and as a result, for a fixed \rmax, the \corrfunc runtime depends weakly on the total number of bins.

To perform the histogram update in the \simd kernel, we use two boolean \simd masks --- a ``unprocessed" mask to identifying the remaining separations, and another ``low" mask for identifying the separations larger than the lower boundary of a specific bin. The separations falling into a given bin are thus uniquely identified by the intersection of these two masks.
The histogram update for that bin is then merely counting the number of pair separations, and amounts to counting the number of bits set in the mask. We have used the explicit hardware instruction, \texttt{popcnt}, for counting the number of set bits. This \texttt{popcnt} operation and the corresponding histogram update are serial operations and constitute a significant fraction of the total runtime.

\subsection{\openmp --- Parallelising for Multi-cores}\label{sec:openmp}
The two biggest challenges in creating an efficient \openmp implementation are -- (i) decomposing the problem into independent tasks and (ii) efficiently using the cache across multiple cores. A correlation function calculation is a ``pleasingly parallel'' problem --- there are multiple ways to partition the entire computation into independent jobs.
A natural division of the overall correlation function computation is in the calculation of all pair-wise separations between a pair of cells. The corresponding \openmp parallelisation can occur at two different scopes -- (i) declaring a \openmp loop over neighbouring cells for any given cell, and (ii) an outer \openmp loop over all cells within the first data-set. The first strategy has better cache utilisation since all \openmp threads share the particle data in the first cell, but the total number of neighbouring cells would limit the thread-scaling. For instance, if there are $3\times3\times3=27$ neighbouring cells, then all requested threads over $27$ would be idle. The second strategy is to create a \openmp loop over primary cells, and calculate pairs for all the associated secondary cells on a single thread. This approach can partition work up to the total number of primary cells ($\gtrsim 1000s$). However, since every thread is working on a different primary cell, the cache utilisation will be worse.

Since there is a strong correlation between the linear cell index and the spatial location, we can improve the cache re-use by ensuring that the \openmp threads process nearby cells. By annotating the \openmp scheduling as \texttt{dynamic}, we ensure that most of the times the various \openmp threads are processing primary cells that are nearby spatially -- thus, increasing the chance the neighbouring cells might get re-used by multiple threads. Since there is only one data-set in an auto-correlation calculation, a secondary cell on a different thread might be the primary cell on another thread. This would lead to better re-use of cached data in auto-correlations compared to cross-correlations.

Thus, all-pairwise computations between the particles in the cell-pair constitutes the minimum amount of work for each \openmp thread. While such a distribution could lead to imbalances in work-load across threads, in practice, we find that the load-imbalance is very small.

To get optimal scaling with threads in multi-core software we need to avoid ``false sharing''. ``False sharing'' occurs when frequently updated variables (on different threads) share the same cache line and causes the entire cache line to be written to memory and then read back. Because memory writes are even slower than memory reads, false sharing can significantly slow down a code. The only memory updates that occur within \corrfunc are to update the pair-counts histogram (and, if requested, the associated average separations and weights) for each bin. Each \openmp thread has a private histogram for the pair counts to avoid ``false sharing", and only this thread-private array is updated within the compute-intensive inner loop (likewise for the average separation and weight arrays).  Once all distance computations for a cell-pair are complete, then these thread-private histograms are added to the global histogram. Note that we assume that pair-counting dominates the overall runtime, and have not implemented parallelism in the domain partitioning step. This assumption may not hold at small particle numbers, where the grid construction can constitute a significant fraction of the total runtime.

\subsection{Summary of optimisations in \corrfunc}
We have discussed a broad range of optimisations implemented in \corrfunc. The
overall optimisation are broadly in two categories -- algorithmic optimisations
to reduce the total number of distance computations, and the software
optimisations that increase the efficiency of such distance
computations. Partitioning the particle domain (\S~\ref{sec:partition} and
\S~\ref{sec:cell_refines}), avoiding duplicate calculations in
auto-correlations (\S~\ref{sec:cell_autocorr}), sorting particles to enable
late-entry and early-exit conditions (\S~\ref{sec:cell_sorting}) are all
examples of algorithmic optimisations. However, the bulk of the novelty in the
\corrfunc package lies in the implemented software optimisations. Storing the
particle positions  contiguously within each cell (\S~\ref{sec:cellstruct}),
explicit vectorisation for calculating pairwise separations
(\S~\ref{sec:cell_simd}), updating the histogram with a \texttt{popcnt} of a
bit-mask are all examples of software
optimisations(\S~\ref{sec:cell_popcnt}). In addition, we have also avoided
executing expensive instructions (\texttt{sqrt}, \texttt{arccos}) wherever
possible\footnote{We have implemented custom approximations for the slow operations --
  \texttt{divisions} (in \texttt{DDrppi\_mocks}) and \texttt{arccos} (in
  \texttt{DDtheta\_mocks}).}.

\section{Benchmarks \& Scaling}\label{sec:benchmarks}
In this section, we present the runtimes and scaling for a different number of particles, \rmax and \openmp threads for each kernel (\fallback, \sse 4.2, and \avx) of \corrfunc.  For comparisons against other codes, see \S~\ref{sec:code_comparison}.  The fiducial catalogue contains $\sim 1.2$ million galaxies on a periodic cube of side 420 \hMpc. Before we delve into the runtime performance for \corrfunc, we will take a step back to examine the expected runtime from a theoretical perspective.

\subsection{Complexity of the Code}
We can now examine the theoretical complexity of the \corrfunc algorithm. Let
\lbox be the side-length of the cube over which \npart points are distributed. Each cell then contains $r_{\mathrm max}^3\times \mathcal{N/V}$ particles. To compute the
correlation function, we first loop over each point, and then {\em all} of
the points in the neighbouring cells -- resulting in a complexity of
$\mathcal{O(NM)}$, where $\mathcal{M} = \mathcal{N}\times\left(\rmax/\lbox\right)^3$. Typical \rmax is $\sim 0.10-0.2\times
\lbox$, therefore ordering the particles in cells of size \rmax should
result in a speedup of $(\rmax/\lbox)^3 \sim 125-1000$ compared to the
brute-force algorithm. When \rmax is comparable to \lbox, the algorithm deteriorates to the brute-force method and tree-based space partitioning approaches might be more suitable~\citep[e.g.,][]{mlpack, feng_FOF_2016}.
For a fixed \rmax, \corrfunc computes the total number of pairs contained
within the cell-pairs, and therefore, directly scales as the number of possible pairs. Therefore, we expect \corrfunc to scale as \onpartsqr with the number of
particles. At fixed \npart, the search volume in \corrfunc scales as \rmaxcubed
and $\thetamax^2$; therefore, we expect the \corrfunc runtime to scale as
$\mathcal{O}(\rmaxcubed)$ and $\mathcal{O}(\thetamax^2)$.

\subsection{Scaling with Number of Particles}\label{sec:npart_scaling}
\begin{figure*}
\includegraphics[width=\figwidth\linewidth,center]{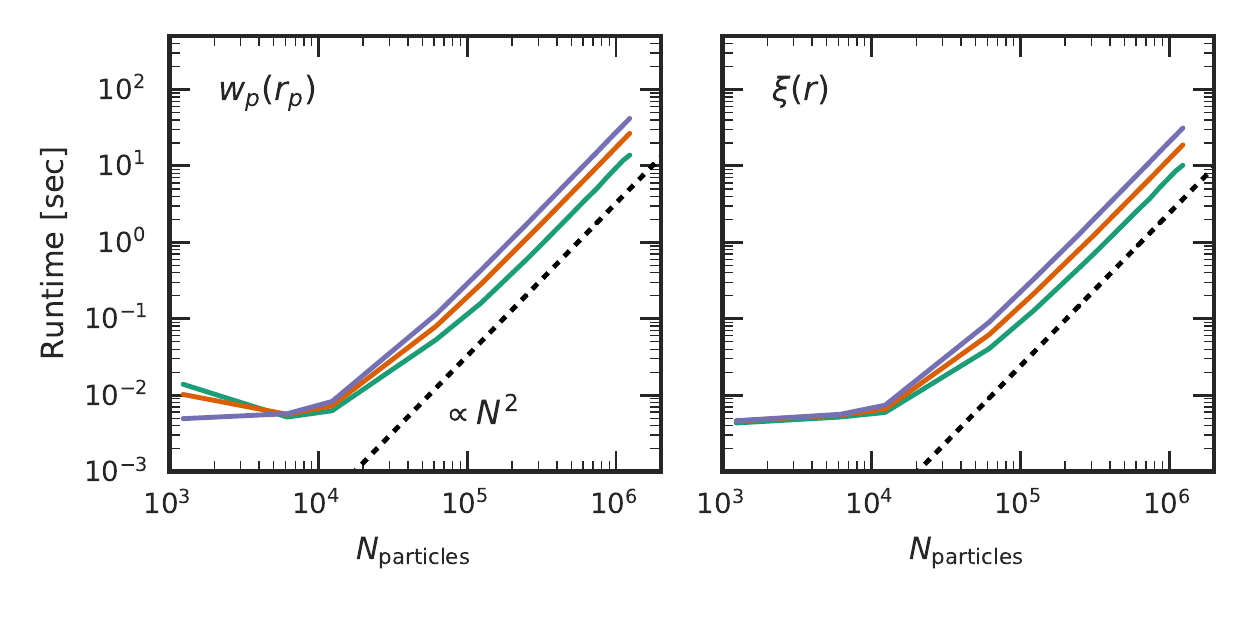}
\includegraphics[width=\figwidth\linewidth,center]{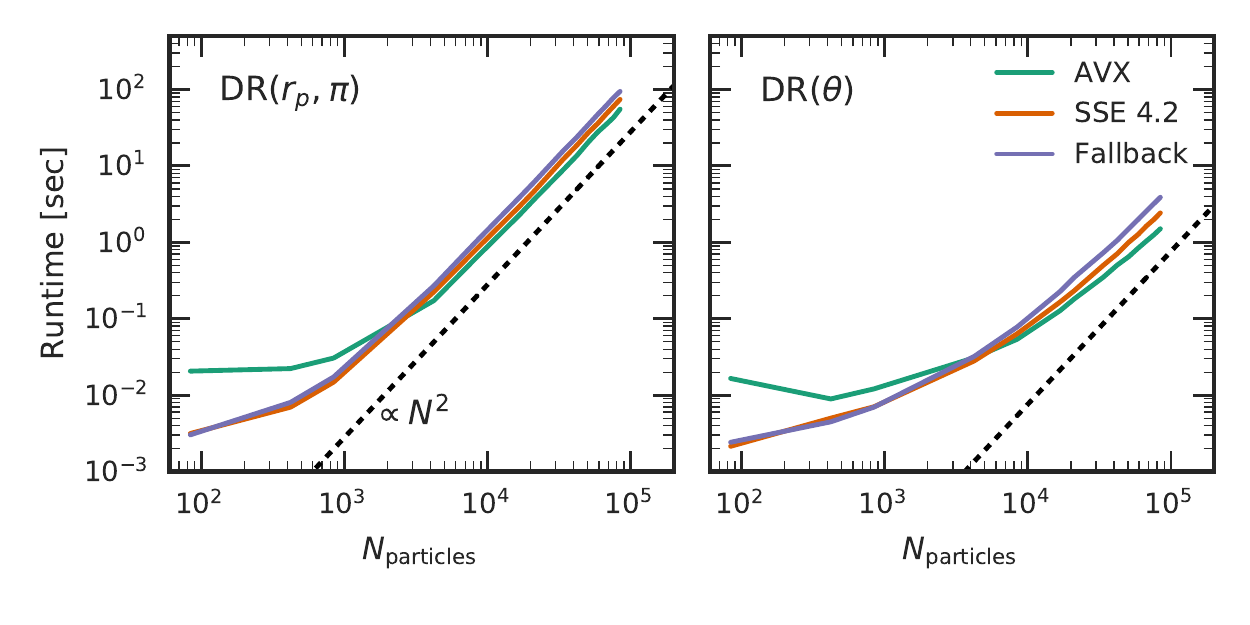}
\caption{Scaling with particle number for (top row) \wprp and \xiofr, and (bottom row) \drrppi and \drtheta.  The first two are auto-correlations on data in a periodic simulation box; the latter two are data-random cross correlations between mock galaxies and randoms with 10 times the number density as the mocks.  The dashed line is a theoretically expected scaling at large \onpart (i.e.~not a fit).}
\label{fig:scaling_numpart}
\end{figure*}

In Fig.~\ref{fig:scaling_numpart}, we show the scaling of four different correlation measures with the number of particles.  In each case, we plot the performance of all three CPU kernels -- the two explicitly vectorised \avx and \sse kernels and the generic \fallback kernel.  To obtain smaller particle sets for scaling tests, we randomly subsampled the fiducial catalogue. We used $\rmax = 84\ \hMpc$ and $\thetamax = 10\degree$ (only for \drtheta).  As expected, we see the runtime scale as \onpartsqr for large \npart. We also see a paradox at low particle numbers for \wprp and \drtheta with less than $10^4$ particles -- the \simd kernels complete faster with increasing particle numbers! We suspect this occurs because, with increasing particle numbers more of the calculation is performed with the efficient \simd instructions, rather than the scalar remainder loop. The gains from the \simd instructions lead to a reduction in the total runtime.

\subsection{Scaling with Maximum Search Radius}\label{sec:rmax_scaling}
\begin{figure*}
\includegraphics[width=\figwidth\linewidth,center]{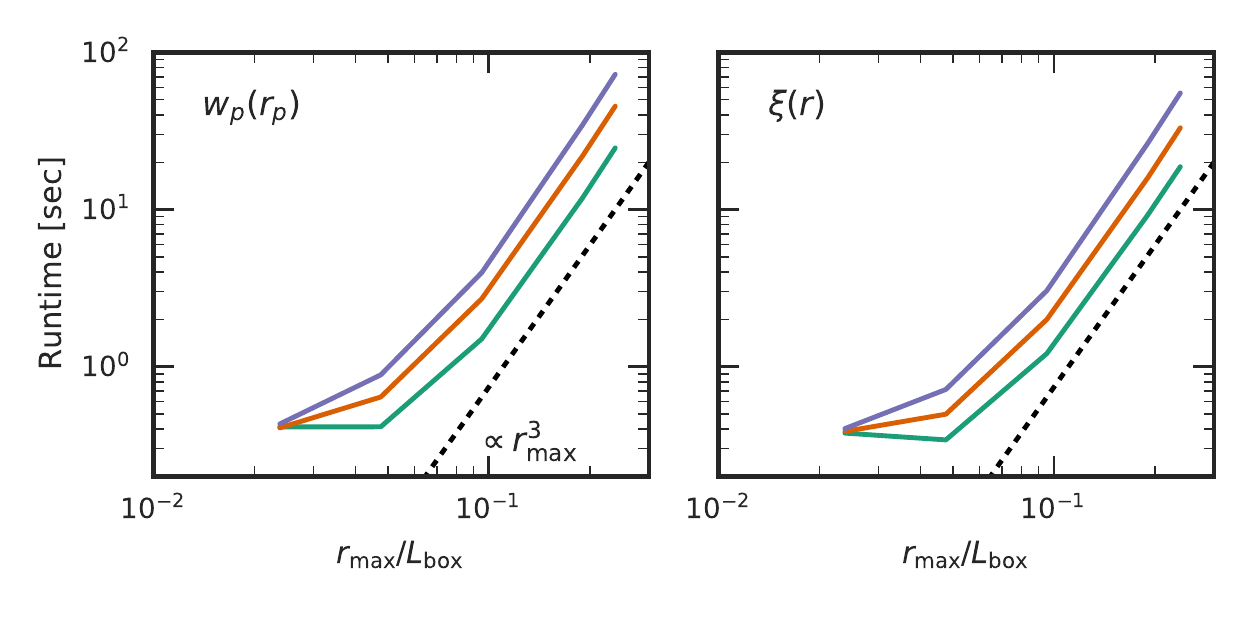}
\includegraphics[width=\figwidth\linewidth,center]{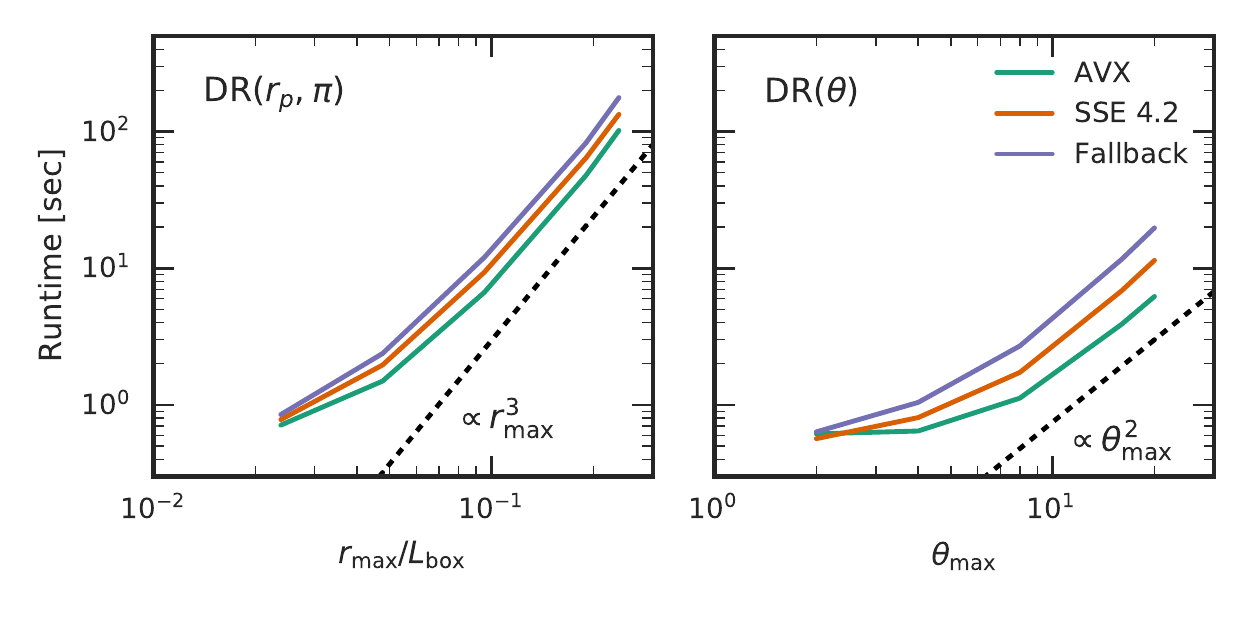}
\caption{Same as Fig.~\ref{fig:scaling_numpart}, but for scaling with \rmax. The \sse and \avx kernels are progressively faster than the generic \fallback kernel.}
\label{fig:scaling_rmax}
\end{figure*}

In Fig.~\ref{fig:scaling_rmax}, we show the scaling for four correlation measures as a function of \rmax, the distance to the outer edge of the last bin.  In the case of \drrppi, we likewise increase $\pi_\mathrm{max}$ so that the expected scaling remains $\mathcal{O}(\rmax^3)$ for large \rmax.  The \drtheta measure scales as $\mathcal{O}(\thetamax^2)$ since the points lie on a 2D surface. In all four cases, we recover the theoretically expected scaling. An interesting feature is for low \rmax ($\sim 0.01-0.05\times\lbox$), we see a flat or decreasing runtime with increasing \rmax. We also saw a similar feature in Fig.~\ref{fig:scaling_numpart}, and we speculate the origins are the same -- at low particle numbers per cell, majority of the computations are performed by the scalar remainder loop. Once the cell occupancy numbers are large enough, the bulk of the computation is done in the efficient \simd instructions, and the total runtime decreases initially. For larger \rmax, most of the calculations are performed by the \simd instructions, and the expected scaling based on the search volume is recovered.

\subsection{Scaling with \openmp threads}\label{sec:openmp_scaling}
\begin{figure*}
\includegraphics[width=\figwidth\linewidth,center]{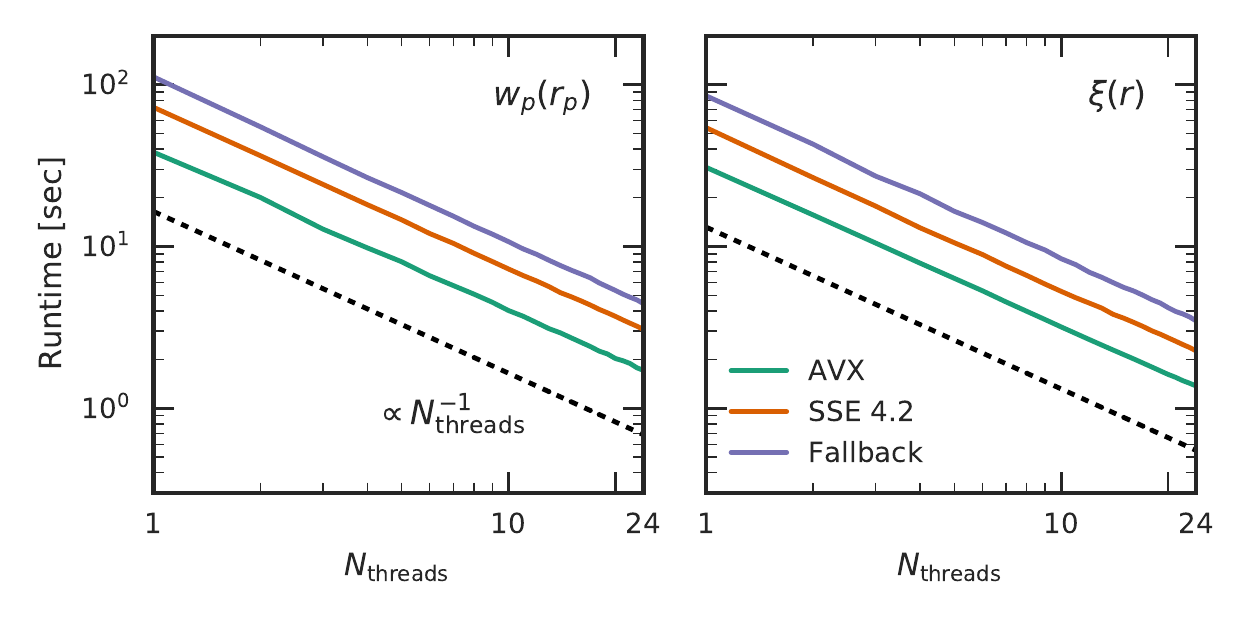}
\includegraphics[width=\figwidth\linewidth,center]{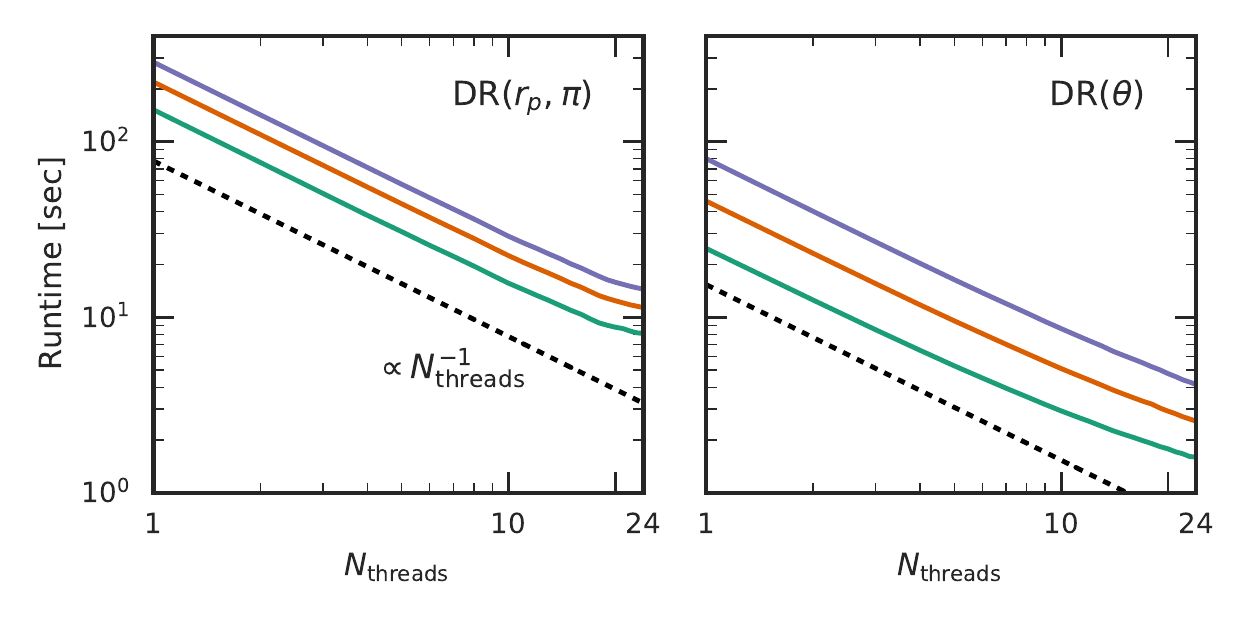}
\caption{Same as Fig.~\ref{fig:scaling_numpart}, but for scaling with number of \openmp threads.  The scaling efficiencies at 24 threads range from 93\% to 105\% for the theory auto-correlations (top row) and 60\% to 80\% for the mock cross-correlations (bottom row)}
\label{fig:scaling_openmp}
\end{figure*}

There are two ways to test how well a software program is parallelised -- i) keep the total problem size fixed regardless of the number of threads (strong scaling) and ii) keep the problem size fixed per thread (weak scaling). Of these two options, the strong scaling test is better at uncovering any performance bottlenecks arising from sub-optimal parallelisation. In Fig.~\ref{fig:scaling_openmp}, we show the results of the strong scaling tests for four pair-counters within \corrfunc. The test platform is the same as in \S\ref{sec:code_comparison} -- a 24-core machine, and the test dataset is the full fiducial mock with $\rmax = 42\ \hMpc$.  As we discussed in \S~\ref{sec:openmp}, the \openmp implementation uses dynamic thread scheduling over primary cells, and all cell-pairs for that primary cell are assigned to a single thread. While the dynamic scheduling ensures that threads are operating on spatially nearby primary cells, the secondary cells in the cell-pair can be more disparate. When the same secondary cell is accessed on multiple threads, then we get the benefits of cached data. For auto-correlations we additionally benefit where the secondary cells on one thread are the primary cell on another thread. In Fig.~\ref{fig:scaling_openmp}, we see that the auto-correlations scale nearly perfectly with $N_\mathrm{threads}$ out to 24 threads (93\% efficiency), while the cross-correlations remain 90\% efficient to $\sim10$ threads and drop to 60--80\% at 24 threads. This drop in efficiency for cross-correlation is a consequence of poorer cache utilisation\footnote{To achieve better cache utilisation, the \openmp loop is now over an array of ``cell-pairs" in \corrfunc \texttt{v2.3.0}.}.

\subsection{Speedup from \simd code}\label{sec:simd_speedup}
The most substantial effort within \corrfunc went towards developing the custom vectorised kernels targeting specific CPU instruction sets. Now that the \corrfunc code-base is mature, we can assess if there are any benefits of explicit vectorisation. \corrfunc \texttt{v2.0.0}, the version presented here, contains three different kernels\footnote{A fourth \avxftf \simd kernel that operates with 512 byte vector registers was added in \corrfunc \texttt{v2.3.0}.} -- one each for \avx, \sse and a generic \fallback kernel. The speedup from perfect vectorisation is directly related to the number of elements processed simultaneously.
Assuming that the compiler generates only scalar instructions for the \fallback kernel, we can immediately see that the maximum possible speedup in the \avx kernels is $8\times$ for \texttt{float32} and $4\times$ for \texttt{float64}. Similarly, for the \sse kernels the maximum possible speedup is $4\times$ for \texttt{float32} and $2\times$ for \texttt{float64}. We can assess the impact of vectorisation on the overall runtimes and then inspect the runtime difference relative to the \fallback kernels as a function of the number of particles belonging to each cell in the cell-pair. Running multiple benchmarks with the \avx and \sse kernels indicates a factor of $\sim 3\times$ and $\sim 2\times$ speedup respectively, relative to the \fallback kernel. While these speedups are at least $2\times$ smaller than the maximum theoretical speedup, the vector kernels are still a significant improvement over the \fallback kernels. Given that we have not reached peak efficiency, it also means that there are potential performance optimisations within the existing vectorised kernels.

\section{Comparison with Other Codes}\label{sec:code_comparison}
Given that performance is one of the design goals of \corrfunc, we need to compare \corrfunc runtimes to external, publicly available correlation function codes. However, we will caution the reader that there are a variety of pitfalls that potentially bias any chosen benchmark (see \url{http://matthewrocklin.com/blog/work/2017/03/09/biased-benchmarks} for a discussion). While we have attempted to be as fair as possible, our unwitting choices may have produced biased benchmarking results.  We encourage the reader (especially correlation function package authors) to do their own benchmarking. The benchmarking code used in this section is publicly available within the \corrfunc repo at \texttt{paper/scripts/generate\_code\_comparison.py}.

\begin{figure}
\centering
\includegraphics[width=\linewidth,center]{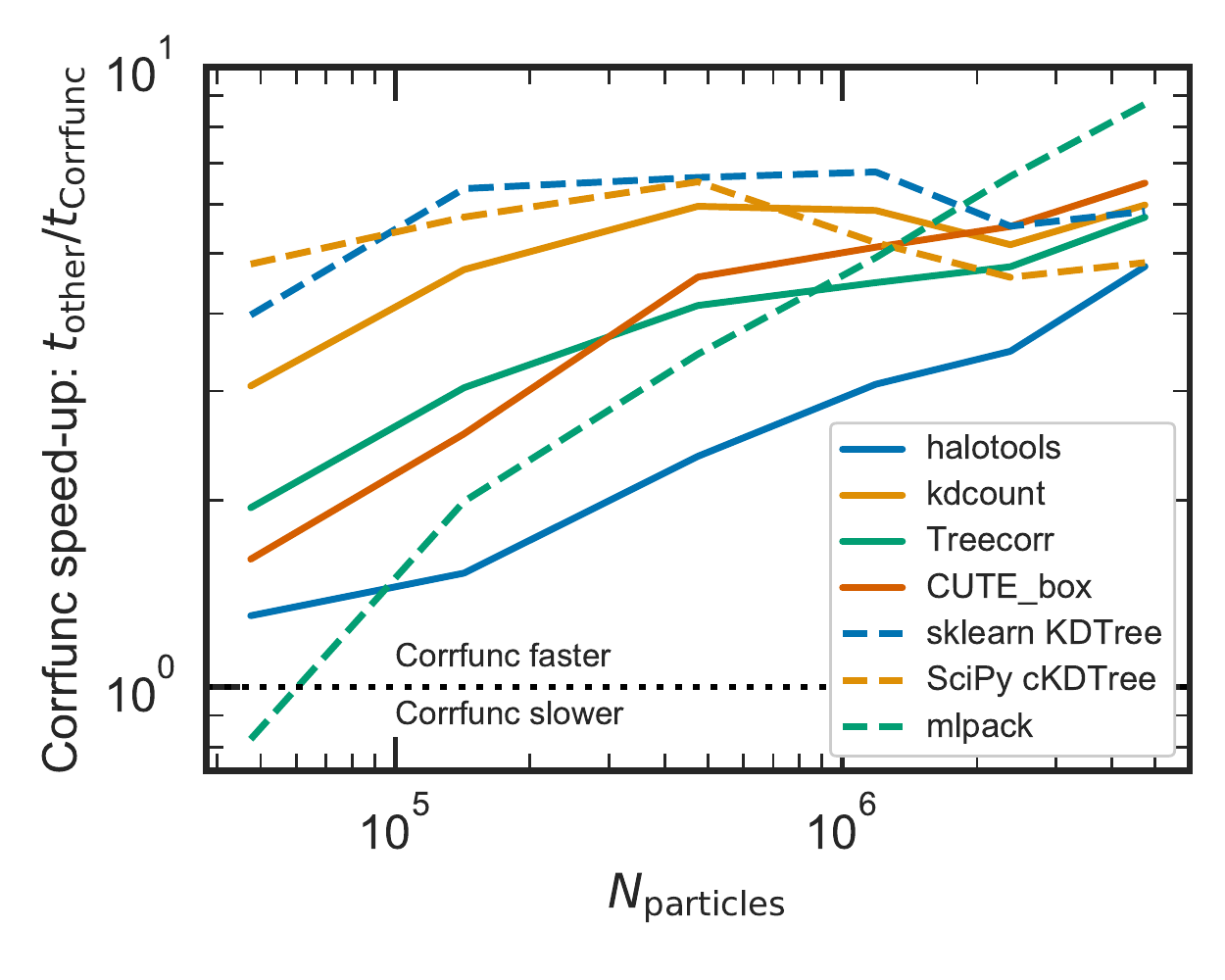}
\caption{\corrfunc runtime vs.~other codes, single-threaded. \corrfunc is faster in the region above the horizontal dashed line. With the exception of \mlpack at low particle numbers, \corrfunc is faster in this benchmark than the other tested codes.}
\label{fig:code_comparison}
\end{figure}

\begin{figure}
\centering
\includegraphics[width=\linewidth,center]{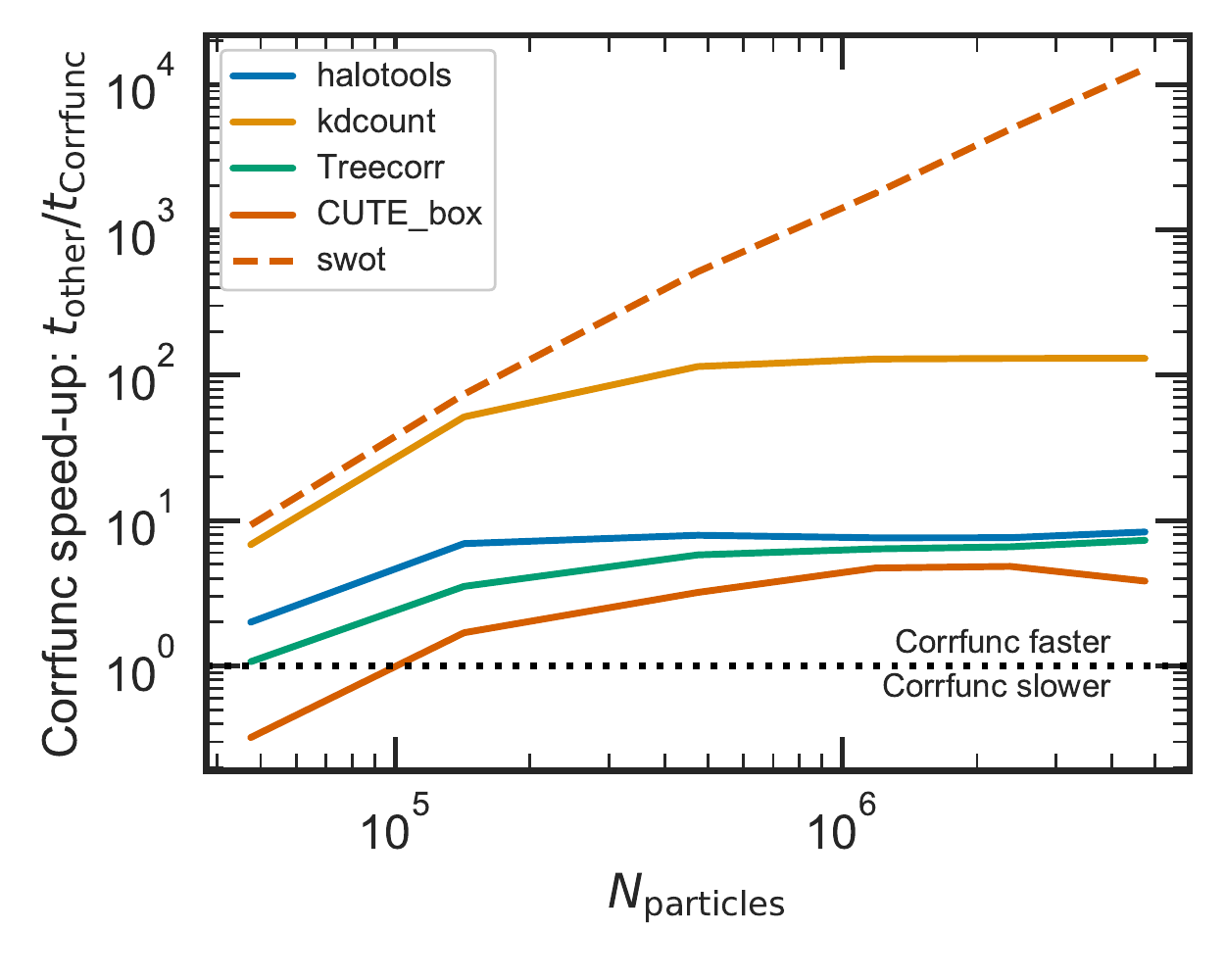}
\caption{\corrfunc runtime vs.~other codes, multi-threaded (all using 24 threads). As with the single-threaded case, \corrfunc is faster in the region above the horizontal dashed line. In this benchmark, \corrfunc was the fastest except at very small particle number (note that the scale of the $y$-axis is different from Fig.~\ref{fig:code_comparison}).}
\label{fig:code_comparison_parallel}
\end{figure}

Our nominal test case is to compute non-periodic pair counts on a clustered dataset in 19 log-spaced bins out to 90 \hMpc\ in a $\sim 1\ h^{-1}\mathrm{Gpc}$ box.  We divide the codes into two categories, serial (Fig.~\ref{fig:code_comparison}) and multi-threaded (Fig.~\ref{fig:code_comparison_parallel}), and run \corrfunc with one or many cores as appropriate.  For each code, we select \corrfunc options that most closely mimics the internal operation of that code.  For example, \treecorr always computes the average pair separation in each bin, so we enable \corrfunc's \texttt{output\_ravg} when comparing against \treecorr.  Such a benchmarking setup means that runtimes of different codes should not be compared against each other in the results that follow; rather, each code's runtime can only be compared against the corresponding \corrfunc runtime.  In all cases, we check that the codes give the same answer (i.e.~same number of binned pair counts).

The clustered dataset was a $z=0.3$ dark matter halo catalogue of 4.7 million objects in a 1100 \hMpc\ box with a lower mass limit of $1.2\times10^{12}\ \hMsun$ from the \textsc{Abacus} project \citep{Garrison+2016}.  Smaller datasets were generated by randomly down-sampling the catalogue.

All timing tests were repeated three times, and the results shown are the average of the three runs.  In general, we have not done any special tuning or optimisation of code parameters for this problem, especially for \corrfunc.  We have also used slower modes of operation for \corrfunc where it is more comparable to the internal operation of the other code, even when a faster mode is available (e.g.~using $\texttt{autocorr} = \texttt{False}$ even when doing auto-pair counts).  In all cases, we have excluded file I/O time and have disabled approximations that reduce pair-count accuracy in exchange for speed.

The test platform was a dual-socket machine with two 12-core Intel E5-2650v4 Broadwell processors at 2.20 GHz with DDR4-2400 RAM.  Turbo-boost and HyperThreading were disabled, and the clock scaling governor was set to \texttt{performance}.  All multi-threaded tests were run with 24 threads.  Absolute \corrfunc times ranged from 0.02~seconds in the fastest multi-threaded case to 500~seconds in the slowest single-threaded case. All calculations were performed in \texttt{float64} precision.

In the following section, we briefly describe each of the eight publicly available 2PCF codes and the comparison methodology.  The script that was used to generate the data for this section is available in the \corrfunc repository as \texttt{paper/scripts/generate\_code\_comparison.py}.

\subsection{\scipy, version 0.18.1 \citep{scipy}}
\begin{itemize}
\item Single-threaded tree code.  Uses a \textit{kd-tree} for spatial sorting, and dual-tree algorithm for pair counting.  Timing includes tree construction and pair counting.
\item All cKDTree options left at defaults (in particular $\texttt{leafsize}=16$).
\item Corrfunc options: $\texttt{autocorr} = \texttt{False}$
\item Corrfunc speed-up: $4.6$ -- $6.5\times$
\end{itemize}

\subsection{\sklearn, version 0.18.1 \citep{sklearn}}
\begin{itemize}
\item Single-threaded tree code.  Uses a \textit{kd-tree}  for spatial sorting, and dual-tree algorithm for pair counting. Timing includes tree construction and pair counting
\item All KDTree options left at defaults (in particular $\texttt{leafsize}=40$), except for specifying dual-tree mode (which was consistently faster in our tests).
\item Corrfunc options: $\texttt{autocorr} = \texttt{False}$
\item Corrfunc speed-up: $4.0$ -- $6.8\times$
\end{itemize}

\subsection{\kdcount, version 0.3.21 \citep{feng_kdcount_2017}}
\begin{itemize}
\item Multi-threaded tree code.  Uses a \textit{kd-tree} for spatial sorting, and dual-tree algorithm for pair counting. Timing includes tree construction and pair counting.
\item All options left at defaults.
\item Corrfunc options: $\texttt{autocorr} = \texttt{False}$
\item Corrfunc speed-up: $3.1$ -- $6.0\times$ (multi-threaded: $6.8$ -- $130.5\times$)
\end{itemize}

\subsection{\halotools, version 0.4 \citep{halotools}}
\begin{itemize}
\item Mesh code that mimics the Corrfunc algorithm in Cython. Divides the domain into rectangular cells. Timing includes mesh construction and pair counting.
\item Fake RR counts were passed to avoid computing RR.  Downsampling was disabled.
\item Corrfunc options: $\texttt{autocorr} = \texttt{False}$
\item Corrfunc speed-up: $1.3$ -- $4.8\times$ (multi-threaded: $2.0$ -- $8.3\times$)
\end{itemize}

\subsection{\treecorr, version 3.3.6 \citep{treecorr}}
\begin{itemize}
\item Multi-threaded tree code.  Uses a ball tree for spatial sorting. Timing includes tree construction and pair counting.
\item \texttt{bin\_slop} was disabled to produce pair counts that exactly agreed with \corrfunc.
\item Corrfunc options: $\texttt{autocorr} = \texttt{True}$, $\texttt{output\_ravg} = \texttt{True}$
\item Corrfunc speed-up: $1.9$ -- $5.7\times$ (multi-threaded: $1.1$ -- $7.3\times$)
\end{itemize}

\subsection{\cutebox, git commit ab33dd8, \citep{cute_alonso_2012}}
\begin{itemize}
\item Mesh code.  Divides the domain into rectangular cells. Timing includes mesh construction and pair counting.
\item Binning changed to linear from log-spaced since \cutebox prefers it.  The internal timers were changed to exclude file I/O time.
\item \cutebox was the only code that returned different pair counts compared to \corrfunc. A small number of pairs seemed to be shifted by one bin.  This could be due to differences between \corrfunc and \cutebox in how floating point math affects decisions about bin boundaries.
\item Corrfunc options: $\texttt{autocorr} = \texttt{True}$, $\texttt{periodic} = \texttt{True}$.
\item Corrfunc speed-up: $1.6$ -- $6.5\times$ (multi-threaded: $0.3$ -- $4.8\times$)
\end{itemize}

\subsection{\mlpack, version 2.0.1 \citep{mlpack}}
\begin{itemize}
\item Single-threaded tree code.  Uses a \textit{kd-tree} for spatial sorting, and dual-tree algorithm for pair counting.  Timing includes tree construction and pair counting.
\item Only supports one bin, and runs out of memory for large numbers of pairs.  \rmax was thus reduced to $36\ \hMpc$ in this test.
\item Timings were recorded using the sum of the reported tree-building and range-search times.  This was significantly faster than the actual runtime, likely because \mlpack explicitly constructs and outputs every pair.
\item Corrfunc options: $\texttt{autocorr} = \texttt{False}$
\item Corrfunc speed-up: $0.8$ -- $8.7\times$
\end{itemize}

\subsection{\swot, version 2.0.1 \citep{swot}}
\begin{itemize}
\item Multi-threaded tree code.  Uses a \textit{kd-tree} for spatial sorting, and dual-tree algorithm for pair counting.  Timing includes tree construction and pair counting
\item As we are only testing exact pair counters in this code comparison, we have used no opening angle approximation.
\item swot only supports evenly spaced bins, and \rmax was reduced by a factor of 100 to make the pair counting faster.  swot was run in \texttt{auto\_3D} mode with re-sampling/covariance and the opening angle approximation disabled.  The data catalogue was also passed as the randoms catalogue.
\item A timer was added to report runtime without including file I/O.  Only the multi-threaded case was tested, as the single-threaded case was prohibitively slow.  The \texttt{Makefile} was modified to include the optimisation flag \texttt{-O3}.
\item Corrfunc was invoked three times on the same data: twice with $\texttt{autocorr} = \texttt{True}$ to emulate DD and RR, and once with $\texttt{autocorr} = \texttt{False}$ to emulate DR.  $\texttt{ouput\_ravg} = \texttt{True}$ was used in all cases.
\item Corrfunc speed-up, multi-threaded: $9.3$ -- $13000\times$
\end{itemize}

In this section, we have a presented both single-threaded and multi-threaded benchmarks against other publicly available correlation function codes. Broadly speaking, \corrfunc is a factor of few faster than all other codes for moderate to high particle loads. There are interesting exceptions where \corrfunc is slower. For instance, \corrfunc at low particle numbers ($\lesssim 10^5$), \mlpack and \cutebox outperforms \corrfunc in the single and multiple threaded tests. For such a low particle load, the absolute runtime is fractions of a second and a degraded \corrfunc performance is unlikely to become a computational bottleneck.

\section{Discussion}\label{sec:discussion}
In this paper, we have presented the highly optimised \corrfunc package for computing correlation functions. Overall, the optimisations presented here can be classified into two broad categories -- (i) improved algorithms and (ii) software co-design to suit the underlying hardware. On the improved algorithm side, \corrfunc partitions the computational domain into 3-D spatial and angular cells (see \S~\ref{sec:partition} and \S~\ref{sec:cell_refines}), and only searches for pairs within neighbouring cells (see \S~\ref{sec:cell_pairs}). Such a partitioning scheme reduces the search volume immensely and reduces the overall number of distance computations. The particles are stored in sorted order within these cells, thereby enabling short-circuits operations based on trigonometric inequalities (see \S~\ref{sec:cell_sorting}). From the software co-design viewpoint, modern CPUs work best when the memory accesses are contiguous and predictable, and when the computations are performed with \simd operations. Within \corrfunc, we duplicate the entire particle distribution to ensure contiguous memory access. Further, we have explicit \simd vector instructions that operate on multiple array elements simultaneously (see \S~\ref{sec:cell_simd}). \corrfunc also avoids expensive operations (like \texttt{sqrt}, or \texttt{arccos}) whenever possible. Computing a correlation function involves an inherently sequential process -- updating a histogram~(see Code~\ref{code:naivecorr}). To minimise the bottleneck from this sequential histogram update, we first create a subset of pairs that {\em fall} into the same bin and simply increment that specific bin once with the total number of pairs~(rather than once each for every pair). Finally, all of the pair-counters presented here are \openmp efficiently parallelised and can scale efficiently on the multi-core modern CPUs.

While \corrfunc is a highly optimised software package, there are still plenty of optimisations left to explore. For instance, we can refine the cells further and carry minimum possible separations along each axis for every cell-pair. With this minimum separation specified, we can alter the late entry and early exit criteria presented in \S~\ref{sec:cell_sorting}\footnote{Implemented in \corrfunc \texttt{v2.3.0} and presented in \citet{corrfunc_exascale_2018}}.

A significant optimisation is possible in cases where only the pair counts are requested for (un-weighted) particles, i.e., no $\langle r \rangle$ within each histogram bin. In such a case, we could maintain a 64-bit integer per histogram bin, and process 64 particle pairs before updating the histogram. Additionally, each histogram bin could be processed independently, and each bit of the all 64 bits in the integer passed to \texttt{popcnt} represent valid data. Currently, the 32-bit integer only represents actual data for the bottom 4 and 8 bits for \texttt{float64} and \texttt{float32} respectively; all the remaining leading bits are identically set to 0. Processing the pairs in this manner would dramatically reduce the total number of calls to \texttt{popcnt}.

Another optimisation requires an update to the algorithm within the vector kernels. As we showed in \S~\ref{section:caches}, retrieving data from memory limits the actual computation rate\footnote{Generally summarised as `Data movement is expensive, flops are free'.}. If we can re-use the retrieved data more effectively, then we get a higher ratio of arithmetic operations per memory retrieval. In the case of \corrfunc, better data re-use could be implemented with a loop-blocking strategy wherein all-pairs are computed for {\em chunks} of $i$ and $j$ particles.
Loop-blocking is a standard optimisation, particularly in matrix multiplications. While not implemented as a matrix multiplication, the \corrfunc algorithm resembles a matrix multiplication, and therefore the \corrfunc runtimes are likely to benefit from a loop-blocking algorithm. However, since the \corrfunc algorithm already reduces the total number of computations with the late-entry and early-exit conditions (see \S~\ref{sec:cell_sorting}), the benefits from loop-blocking might be more limited. Our preliminary attempts at implementing loop-blocking produced significant slow-downs for low to moderate particle numbers, and we plan to revisit in the future. Another advantage of loop-blocking is that the kernels then process in constant chunks of $i$ and $j$ particles, rather than actual (variable) cell occupancy numbers. Since these constant chunks are known at compile-time, the compiler can generate more optimised code -- leading to performance benefits beyond the data re-use.

Finally, the correlation function code resembles a matrix multiplication with a high ratio of arithmetic operations to memory access. Matrix multiplication algorithms are known to benefit from loop-blocking, and loop-unrolling, as well as offloading to a GPU. There are a wide-variety of highly optimised Basic Linear Algebra Subprograms (BLAS) libraries, as well as GPU BLAS libraries (e.g., \texttt{cuBLAS}) that can be utilised the efficiently calculate the pair-wise separations at the cell-pair level.

One major challenge for the \corrfunc package is the number of duplicate lines of source code. Each pair-counter performs two unique operations -- (i) to compute the pair-separation and (ii) to compute the histogram bin index. Together, these two lines require $\sim 50$ lines of code. However, a large amount of boilerplate code is necessary to set up the domain partitions, associating pairs of cells, offloading to the appropriate \simd kernel, and finally collecting the results across multiple threads. These operations are common for all pair-counters and result in a lot of duplicated source code. Duplicated code requires unnecessary maintenance overheads, and can easily be a source of bugs. To improve the sustainability of the \corrfunc and guided by our experience in developing \corrfunc, we plan to evaluate an auto-generated code-base in future. An auto-generated code-base will allow us to implement custom kernels to tackle a broad range of user-cases, {\em without} increasing the actual lines of maintained source code.

\section{Conclusions}\label{sec:conclusion}
We have presented a suite of three blazing-fast correlation function codes within the \corrfunc software package.  At a high level, \corrfunc achieves its performance from the following aspects:
\begin{itemize}
\item Domain knowledge --- The typical correlation function calculation only requires pairs over a much smaller scale (\rmax) compared the spatial extent of the dataset ($\mathcal{L}$). By constructing 3-D spatial grids, and only computing distances for point pairs that can be within \rmax, \corrfunc trims the search volume significantly.
\item Cache locality --- Memory access is typically much slower compared to the CPU speeds. By arranging the particles contiguously within each cell of the 3D grid, \corrfunc significantly improves the memory access speeds.
\item Vectorisation--- Modern CPUs contain wide vector registers that can process multiple elements simultaneously. Even after experimenting with a variety of different formulations, the compilers so far have not been able to generate a vectorised code for a correlation function. By using explicit vector intrinsics, \corrfunc computes multiple pairs simultaneously.
\item Multicore algorithm --- Calculating a correlation function is inherently a parallel problem. \corrfunc uses \openmp parallelisation without any shared mutable resource. In our tests, \corrfunc shows excellent strong scaling characteristics.
\end{itemize}

High-performance and user-friendly research software like \corrfunc is required for modern research. However, designing, writing, and maintaining such a software package is quite time-intensive. All of the \corrfunc algorithm and associated source code have been conceived and have evolved over more than five years. The authors of \corrfunc would like to thank the \corrfunc users for citing the package through the \texttt{https://ascl.net/1703.003} entry prior to this paper being published.

\section*{Acknowledgements}
The authors would like to thank the referee for constructive comments that helped improve the clarity of the paper. MS would like to thank A. Berlind, J. Piscionere, B. Wibking, Q. Mao and A. Hearin for constructive discussion about \corrfunc over the years. MS would particularly like to thank J. Piscionere for significantly improving the user experience by crashing \corrfunc in novel ways. MS was primarily supported by NSF Career Award (AST-1151650) during \corrfunc design and development. MS was also supported by the Australian Research Council Laureate Fellowship (FL110100072) awarded to Stuart Wyithe and by funds for the Theoretical Astrophysical Observatory (TAO). TAO is part of the All-Sky Virtual Observatory and is funded and supported by Astronomy Australia Limited, Swinburne University of Technology, and the Australian Government.  The latter is provided though the Commonwealth's Education Investment Fund and National Collaborative Research Infrastructure Strategy (NCRIS), particularly the National eResearch Collaboration Tools and Resources (NeCTAR) project. Parts of this research were conducted by the Australian Research Council Centre of Excellence for All Sky Astrophysics in 3 Dimensions (ASTRO 3D), through project number CE170100013. This research has made use of NASA's Astrophysics Data System, the arXiv.org preprint server, and extensive use of the Intel Intrinsics Guide~( \url{https://software.intel.com/sites/landingpage/IntrinsicsGuide/}). This research has used \texttt{sglib}~\citep{vittekBorovanskyMoreauTurin2006}, \texttt{python}~(\url{https://www.python.org/}),
\texttt{numpy}~\citep{van2011numpy}, \texttt{matplotlib}~\citep{Hunter:2007}, \texttt{GSL}~(\url{http://www.gnu.org/software/gsl/}),
and \texttt{The (Astronomy) Acknowledgment Generator}~(\url{http://astrofrog.github.io/acknowledgment-generator/}).

\bibliographystyle{mnras}
\bibliography{master}

\appendix
\section{Conventions for Separations in Correlation Functions}\label{sec:appendix_separations}
\subsection{Pair-counting in Cartesian volumes}\label{sec:cartesian_npairs}
For Cartesian volumes (simulation boxes), the separation between points is the standard Euclidean separation:
\begin{align}
\begin{split}
r^2 &= (x_1 - x_2)^2 + (y_1 - y_2)^2 + (z_1 - z_2)^2, \\
r_p^2 &= (x_1 - x_2)^2 + (y_1 - y_2)^2, \\
\pi &= \left|z_1 - z_2\right|.
\end{split}
\label{eqn:sep}
\end{align}
Generally speaking, \corrfunc accepts spatial bins defined by the first two columns of a text file. This text file is specified through a filename parameter on the command-line, or equivalently via an \texttt{numpy} array through the Python interface. We will refer to this input file as \texttt{file\_with\_bins} for the remainder of the paper. We always assume that the $z$ direction is the line-of-sight, i.e., the $\pi$ direction, and we assume any relevant redshift-space distortions have already been imposed on the particle positions. There are four distinct correlation functions defined for Cartesian volumes, and we will outline what each one of the correlation functions assumes for the binning.
\begin{itemize}
\item $\xir$ and $\xi(r)$ - The separation is $r$ and calculated in full 3-D space. Bins in the first two columns of \texttt{file\_with\_bins} are assumed to specify $r$
\item $\ddrppi$ - The separation is $r_p$ and $\pi$ as shown in Eqn.~\ref{eqn:sep}. Bins in the first two columns of \texttt{file\_with\_bins} are assumed to specify $r_p$. Bins in $\pi$ are of width $1$, linearly spaced between $[0, \pimax]$
\item $\wprp$ - The separation is $r_p$ as shown in Eqn.~\ref{eqn:sep}. Bins in the first two columns of \texttt{file\_with\_bins} are assumed to specify $r_p$. All points with separation up to $\pimax$ in the $\pi$ direction are included.
\end{itemize}
\subsection{Pair-counting in Spherical volumes}\label{sec:spherical_npairs}
\begin{figure}
\centering
\begin{tikzpicture}[scale=0.6, every node/.style={transform shape}]
\tikzset{>=latex}
\tikzset{
    >=triangle 45,
}
  \node (o) at (0,0) {};
  \node (a) at (120:5cm) {};
  \node (b) at (60:8cm) {};

  \draw[very thick, ->] (o) -- (a) node[pos=0.5, sloped, below] {\Large \vvec{v_1}};
  \draw[very thick, ->] (o) -- (b) node[pos=0.5, sloped, below] {\Large \vvec{v_2}};

  \draw [dashed, thick, ->] (b) -- coordinate[midway](l) (a) node[pos=0.5, sloped, above] {\Large \vvec{s} = \vvec{v_1} - \vvec{v_2}};

  \draw [dashed, thick, ->] (o) -- (l) node [pos=0.7, below, sloped] {\Large $\vvec{\mathbf{\ell}}$};


  \draw [dashed, thick, ->] (a) -- ($(a) + (l)$) coordinate (apl) node [pos=0.5, sloped, above] {\Large $\vvec{\ell}$ = $\dfrac{1}{2} \left(\vvec{v_1} + \vvec{v_2} \right)$} ;

  \coordinate (p) at ($(a)!(b)!(apl)$) {};

  \draw [thick] ($(a)!(b)!(apl)$) -- (b) node [pos=0.5, sloped, above]{\huge $\mathbf{r_p}$};

  \draw [thick] (a) -- ($(a)!(b)!(apl)$) node [pos=0.5, sloped, below] {\Huge $\mathbf{\pi}$};

  \MarkRightAngle{a}{$(a)!(b)!(apl)$}{b};

  \draw
    pic["{\Large {$\Delta \mathbf{\sigma}$}}", thick, draw=black, anchor=east, <->, angle eccentricity=1.2, angle radius=2cm]
    {angle=b--o--a};

    \eye{1}{0}{-1}{90}
\end{tikzpicture}
\caption{\small We follow the conventions of \citep{fisher_1994_rp_pi} to define the separations for points on the sky. }
\label{fig:sky_sep_defn}
\end{figure}
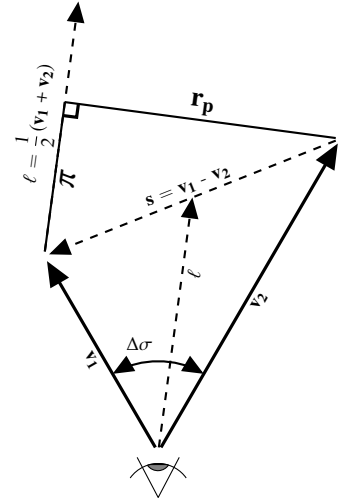

When working with galaxies with positions defined in spherical coordinates (i.e., on sky positions), we follow the conventions in \citep{fisher_1994_rp_pi}. We define the line-of-sight vector ($\vvec{\mathbf{\ell}}$ in Fig.~\ref{fig:sky_sep_defn}) as the line connecting the observer to the mid-point of the line joining the two points.

\subsubsection{\texorpdfstring{\xirppi}{xi(rp, pi)}}
For defining projected correlation functions on the sky, we use the following equations~\citep{fisher_1994_rp_pi}:
\begin{align}
\begin{split}
\vvec{s} &= \vvec{v_1} - \vvec{v_2}\\
\vvec{\ell} &= \dfrac{1}{2} \left( \vvec{v_1} +  \vvec{v_2} \right), \\
\pi &=  \vvec{s} \cdot \vvec{\ell}/ \norm{\vvec{\ell}}, \\
r_p^2 &=  \vvec{s} \cdot \vvec{s} - \pi^2,\\
\end{split}
\end{align}
Bins in the first two columns of \texttt{file\_with\_bins} are assumed to specify $r_p$. Bins in $\pi$ are of width $1$, linearly spaced between $[0, \pimax]$.

\subsubsection{\wtheta}
For the angular correlation function, \wtheta, the separation is the angle between the two vectors, $\vvec{v_1}$ and $\vvec{v_2}$, and is represented in Fig.~\ref{fig:sky_sep_defn} by $\Delta\sigma$. With vector dot product, we can calculate the angular separation using the following equations:
\begin{align}
\begin{split}
\cos \Delta\sigma  &= \dfrac{\vvec{v_1} \cdot \vvec{v_2}}{\norm{\vvec{v_1}}
  \norm{\vvec{v_2}}}, \\
 &= \vvec{v_1} \cdot \vvec{v_2}, \quad \because \norm{\vvec{v_1}} = \norm{\vvec{v_2}}=1, \\
\Delta\sigma &= \arccos(\vvec{v_1} \cdot \vvec{v_2}).
\end{split}
\end{align}
\corrfunc only tackles angular separations between $0\deg$ and $180\deg$.

\section{Pair-counts with 0 as minimum separation}\label{appendix:self-pairs}
In \corrfunc, we aim to return identical results an auto-correlation of a data-set and the cross-correlation of the data-set with itself.  There are a few minor modifications necessary to the pair-counting behaviour to ensure this.  First, the DD term in auto-correlations must include the self-pair if the first bin starts at zero separation.  This is because the cross-correlation will always consider zero-separation particles, and the auto-correlation should mimic the cross-correlation.  We do not explicitly enumerate and count these pairs in the auto-correlation, but instead, add $N$ to the appropriate bin after pair counting is complete.

Second, the RR term must be adjusted to use a particle density of $(N-1)/V$ instead of $N/V$.  This adjustment is necessary because the primary particle is never available at a non-zero distance to make a pair with, leaving $N-1$ particles spread throughout the rest of the volume.  Since every particle gets a turn being the primary, the expected RR pair counts in bin $i$ with volume $\Delta V_i$ become:
\begin{align}
RR_i &= \Delta V_i N \rho \\
     &= \Delta V_i N \frac{N-1}{V}
\end{align}

Using a density of $N/V$ leads to a biased estimator, in that $\xi(r)$ of a uniform random set of particles will not yield 0, but instead $-1/N$.

Finally, for consistency with the DD behaviour, the RR term must include self-pairs if the first bin includes 0.


\bsp	
\label{lastpage}
\end{document}